\documentclass[%
 reprint,
 amsmath,amssymb,
 aps,
 floatfix
]{revtex4-2}

\usepackage{graphicx}
\usepackage{dcolumn}
\usepackage{bm}

\begin{document}

\preprint{APS/123-QED}

\title{Analytic approximations for the primordial power spectrum with Israel junction conditions}

\author{D.D. Dineen}
\email{ddd25@cam.ac.uk}

\author{W.J. Handley}
\email{wh260@cam.ac.uk}

\affiliation{Astrophysics Group, Cavendish Laboratory, J.J. Thomson Avenue, Cambridge CB3 0HE, United Kingdom}
\affiliation{Kavli Institute for Cosmology, Madingley Road, Cambridge, CB3 0HA, United Kingdom}

\date{March 18, 2024}
\begin{abstract}
    This work compares cosmological matching conditions used in approximating generic preinflationary phases of the Universe. We show that the joining conditions for primordial scalar perturbations assumed by \citet{2003JCAP...07..002C} are inconsistent with the physically motivated Israel junction conditions; however, performing general relativistic matching with the aforementioned constraints results in unrealistic primordial power spectra. Eliminating the need for ambiguous matching, we look at an alternative semianalytic model for producing the primordial power spectrum allowing for finite duration cosmological phase transitions. 
\end{abstract}

\maketitle


\section{Introduction}
The standard model of cosmology consists of a Universe filled with cold dark matter and a cosmological constant known as $\Lambda$CDM \cite{2011ApJS..192...18K, 2014A&A...571A..22P, 2020A&A...641A..10P}. Inflation is a period of exponential expansion in the very early Universe which is an additional ingredient to the current paradigm that solves several issues of standard big bang cosmology such as the cosmological horizon, flatness, and monopole problems \cite{1981PhRvD..23..347G, 1979JETPL..30..682S, 1982PhLB..108..389L}. Most notably, inflation provides a causal theory of structure formation whereby quantum fluctuations deep inside the comoving horizon grow to macroscopic scales with the accelerated expansion of the Universe \cite{2009arXiv0907.5424B,1992PhR...215..203M}. The primordial power spectrum provides a statistical measure of these scalar fluctuations and is found to be nearly scale invariant by current observations \cite{2014A&A...571A..22P, 2020A&A...641A..10P}. 

We consider a cosmological scenario in which the Universe evolves from the initial singularity into a noninflating state, termed kinetic dominance, where the potential energy of the inflaton is exceeded by its kinetic energy  \cite{2003JCAP...07..002C, 2014PhRvD..89f3505H, 2016PhRvD..94b4041H, 2019PhRvD.100b3501H, 2019PhRvD.100b3502H}. Generating a power spectrum of scalar primordial perturbations generally requires numerical solutions to the equations describing the background evolution of the Universe which in turn demands a choice of the functional form of the inflationary potential. The ability to produce primordial power spectra which do not require a selection of inflationary potential is useful in that it allows for generic analyses of the early Universe \cite{2021PhRvD.104f3532G, 2021PhRvD.103l3513H,2021PhRvD.103b3519T}. \citet{2003JCAP...07..002C} present a model of this form wherein the background Universe is approximated by an instantaneous transition between a primordial phase of kinetic dominance and inflation.

Our focus will be on formulating physically acceptable matching conditions which join scalar perturbations across cosmological phase transitions defined by a jump in the equation of state of the scalar field. We are concerned both with the primordial power spectrum for the cosmological scenario of interest and coming to general conclusions as to the effects of instantaneous phase transition on the primordial power spectrum. A theory on the propagation of primordial perturbations through a cosmological transition is present in the literature with application to three scenarios. These are, transitions between inflation and a slow-roll violating phase
\cite{1995PhRvD..52.5549D, 2008PhRvD..77h3501W, 2012JCAP...12..012C, 2012JCAP...12..018N, 2016PhRvD..93l3519A,2019JCAP...06..028B,2020JCAP...04..048O,2021PhRvD.103b3535T,2023arXiv230809273C},
the change from contraction to expansion in an inflationary alternative known as a bouncing universe \cite{2001JHEP...11..056B,2002PhLB..524....1L,2002PhRvD..65l3513M,2002PhLB..526..173L,2002PhRvD..65j3522F,2007JCAP...06..014C}, and finally, to evolve the primordial power spectrum to current observations, the transition between inflation and reheating is considered \cite{PhysRevD.43.3802, 1998PhRvD..57.3302M,2009PhRvD..80l3526A, 2019JCAP...12..018C, 2020PhRvD.101d3511D}. These references provide a starting point for the novel analysis contained in this work which applies Israel junction conditions to the matching of primordial scalar perturbations in the Contaldi approximation.

We subsequently introduce an alternative model which smoothly joins the analytic scaling of the comoving horizon for a phase of kinetic dominance preceding inflation, which can be used to generate the primordial power spectrum from finite duration cosmological phase transitions. Power spectra produced from arbitrarily sudden cosmological phase transitions will prove fruitful in comparing to those arising from instantaneous transitions in the Contaldi approximation. Although this method does not demand a choice in functional form of the inflationary potential, the Hamilton-Jacobi formalism presents the opportunity for phenomenological study.

This paper is organized as follows. Section II details theoretical background and establishes notation for gauge invariant variables and the primordial power spectrum. In Sec. III the Contaldi approximation is introduced as a potential independent method for producing an analytic primordial power spectrum. Sec. IV proposes the use of Israel junction conditions to derive cosmological matching conditions for primordial scalar perturbations. In Sec. V the primordial power spectra produced from applying cosmological matching conditions to the Contaldi approximation are shown. Sec. VI presents an alternative model for generating the primordial power spectrum from a smooth analytic background. Conclusions and directions for future work are presented in Sec. VII.

\section{Background}

Cosmic time derivatives, d$t$, will be represented by overdots and conformal time derivatives, d$\tau$,  by primes unless otherwise specified. As well $V_{,\phi}=\frac{dV}{d\phi}$ and partial derivatives are denoted by commas. All equations are given in natural units such that $c = \hbar = 8\pi G=1$. We work in the case of a flat universe where the curvature of background space is $K=0$. The metric signature used is $(+, -, -,-)$.

The background theory developed in this section uses Refs. \cite{2009arXiv0907.5424B,1992PhR...215..203M} unless otherwise stated. 

\subsection{Single-field inflation}

The simplest models of inflation involve a single scalar field, $\phi$, known as the inflaton, whose self-interactions are characterized by the inflationary potential, $V(\phi)$. The action is composed of the summation of the Einstein-Hilbert action and the action of a scalar field with a canonical kinetic term, 

\begin{equation}
    S=\int d^4x\sqrt{|g|}\Big(\frac{1}{2}R+\frac{1}{2}\nabla^{\mu}\phi \nabla_{\mu}\phi-V(\phi) \Big),
\end{equation}

\noindent
where $g_{\mu\nu}$ is the metric and $R$ is the Ricci scalar. Under the assumptions of the cosmological principle of homogeneity and isotropy the Friedmann-Robertson-Walker (FRW) metric is utilized,

\begin{equation}
    ds^2=dt^2-a(t)^2\delta_{ij}dx_idx_j.
\end{equation}

 Using the stress-energy tensor, $T_{\mu \nu}$, for a perfect fluid in thermodynamic equilibrium and applying the FRW metric to the Einstein field equations, the Fridemann and Klein-Gordon equations can be obtained which comprise the background expressions governing the dynamics of the geometry and evolution of the scalar field. These are Eqs. (3)-(4) and Eq. (5), respectively,

\begin{equation}
    H^2=\frac{1}{3}\Big(\frac{1}{2}\dot{\phi}^2+V(\phi)\Big),
\end{equation}

\begin{equation}
    \dot{H}+H^2=-\frac{1}{3}\Big(\dot{\phi}^2-V(\phi)\Big),
\end{equation}

\begin{equation}
    \ddot{\phi}+3H\dot{\phi}+V_{,\phi}=0,
\end{equation}

\noindent
where $a$ is the scale factor and $H=\frac{\dot{a}}{a}$ is the Hubble parameter. Specifying initial conditions on $\phi$, $\dot{\phi}$ and knowing the form of the scalar field potential, Eqs. (3) and (5) can be solved to fully specify the evolution of a flat universe.

\subsection{Mukhanov-Sasaki equation}

The early Universe was very nearly homogeneous; therefore, it is sufficient to consider linear perturbations of the scalar field about its homogeneous background, 

\begin{equation}
    \phi(t,\mathbf{x})=\bar{\phi}(t)+\delta\phi(t,\mathbf{x}),
\end{equation}

\noindent
and linear perturbations of the metric about its background,

\begin{equation}
   g_{\mu\nu}(t,\mathbf{x})=\bar{g}_{\mu\nu}(t)+\delta g_{\mu\nu}(t,\mathbf{x}).
\end{equation}

\noindent
In real space, the scalar vector tensor (SVT) decomposition of the metric perturbations is

\begin{multline}
     ds^2=(1+2\Phi)dt^2
     +2a(t)(\partial_i B-S_i)dx_idt\\
     -a(t)^2(1-2\Psi)\delta_{ij}dx_idx_j\\
     -a(t)^2(2\partial_i \partial_jE
     +2\partial_{(i}F_{j)}+h_{ij})dx_idx_j.
\end{multline}

\noindent
In the case of linear perturbations, scalar, vector, and tensor components do not dynamically mix, and hence, we can neglect vector and tensor perturbations in the following derivations. Threading and slicing of perturbed spacetime is not unique, and thus, it is useful to define a gauge invariant combination of the scalar type metric and scalar field perturbations to ensure fluctuations cannot be removed by a coordinate transformation. The comoving curvature perturbation, $\mathcal{R}$, is defined as

\begin{equation}
    \mathcal{R} \equiv \Psi-\frac{H}{\dot{\bar{\phi}}}\delta \phi,
\end{equation}

\noindent
which can be geometrically interpreted as a measure of the spatial curvature of comoving or constant scalar field value ($\delta \phi=0$) hypersurfaces. 

Using Eq. (6) for the perturbed scalar field and Eq. (9) for the gauge invariant comoving curvature perturbation, the action to second-order in $\mathcal{R}$ is

\begin{equation}
    S_{\mathcal{R}}=\frac{1}{2}\int d^4 x a^3 \Big(\frac{\dot{\phi}}{H}\Big)^2\Bigg(\dot{\mathcal{R}}^2-\Big(\frac{\partial_i\mathcal{R}}{a}\Big)^2\Bigg),
\end{equation}

\noindent
where the Mukhanov variable, $v$, is assigned as

\begin{equation}
    v \equiv z\mathcal{R},
\end{equation}

\noindent
specifying $z$ as the following function of cosmic time:

\begin{equation}
    z \equiv \frac{a\dot{\phi}}{H}.
\end{equation}

\noindent
Changing to conformal time, $\tau=\int\frac{dt}{a}$, gives

\begin{equation}
    S_v=\frac{1}{2}\int d^3 x d\tau\Big((v')^2-(\partial_iv)^2+\frac{z''}{z}v^2\Big).
\end{equation}

\noindent
Taking the Fourier transform in spatial coordinates followed by extremizing the resulting action gives the Mukhanov-Sasaki (MS) equation in terms of the Mukhanov variable and derivatives with respect to conformal time,

\begin{equation}
    v_{k}''+\Big(k^2-\frac{z''}{z}\Big)v_{k}=0.
\end{equation}

\noindent
This equation in the form of a simple harmonic oscillator with time dependent mass, $\frac{z''}{z}$, describes the evolution of comoving curvature perturbations with comoving wave number $k$. Solutions to the MS equation in general require numeric integration of the coupled background expressions (3) and (5) in order to determine the evolution of the scale factor and Hubble parameter which control the behavior of $\frac{z''}{z}$.

\subsection{Primordial power spectrum}
The primordial power spectrum is the Fourier transform of the two-point function of comoving curvature perturbations. This is 

\begin{equation}
    P_{\mathcal{R}}(k)=\Big|\frac{v_k}{z}\Big|^2.
\end{equation}

\noindent
Recalling Eq. (11) for the Mukhanov variable in terms of the comoving curvature perturbation, the dimensionless power spectrum can be written as

\begin{equation}
    \mathcal{P_R}(k)=\lim_{k\ll aH}\frac{k^3}{2\pi^2}|\mathcal{R}_{k}|^2,
\end{equation}

\noindent
where the limit ${k\ll aH}$ indicates that modes are evaluated upon exiting the comoving horizon and freezing-out. 

In the six parameter $\Lambda$CDM cosmology the primordial power spectrum is parametrized by $n_s$, the scalar spectral index and $A_s$, the amplitude of fluctuations, through the following power law:
\begin{equation}
    \mathcal{P_{R}}(k)=A_s\Big(\frac{k}{k_*}\Big)^{n_s-1}.
\end{equation}

\noindent
Here, $k_*$ is an arbitrary reference scale referred to as the pivot scale which sets the location of the cutoff in the power spectrum \cite{2020A&A...641A..10P}.

\section{Contaldi approximation}

As previously mentioned, the MS equation defined by (14) has in general no analytic solutions; however, analytic primordial mode functions can be obtained in a number of special cases such as when the approximation $\frac{z''}{z}\approx\frac{a''}{a}$ holds \cite{2003JCAP...07..002C}. 

The first slow-roll parameter, $\varepsilon$, in terms of the scalar field and cosmic time derivatives is \cite{2009arXiv0907.5424B}

\begin{equation}
    \varepsilon=\frac{1}{2}\frac{\dot{\phi}^2}{H^2}.
\end{equation}

\noindent
Using the above expression and recalling Eq. (12), we can equivalently define $z$ as 

\begin{equation}
    z=\pm a\sqrt{2\varepsilon},
\end{equation}

\noindent
where working with the positive root corresponding to the choice that $\dot{\phi}>0$. From Eq. (19), it is clear that $z\propto a$ and thus, $\frac{z''}{z}\approx\frac{a''}{a}$ when $\varepsilon$ is constant in time. When these conditions hold, the following expression may be taken as an approximation for the perturbation evolution equation:

\begin{equation}
    v''_k+\Big(k^2-\frac{a''}{a}\Big)v_k=0.
\end{equation}

\noindent
This MS equation approximation has analytic solutions during kinetic dominance and inflation which will be used to define the analytic primordial power spectrum in the Contaldi approximation \cite{2003JCAP...07..002C}.

Figure 1 shows the analytic evolution of the background in the Contaldi approximation with a comoving horizon which instantaneous transitions between an era of kinetic dominance and de Sitter inflation. We set for mathematical convenience the transition to be at $\tau=0$. The comoving horizon is matched at the transition taking the value of $\frac{1}{k_{\mathrm{t}}}\equiv\frac{1}{a_{\mathrm{t}}H_{\mathrm{t}}}$ as $a$ and $H$ are required to be matched in this model \cite{2003JCAP...07..002C}. The Contaldi approximation demands a jump in the first-slow roll parameter, $\varepsilon$, and equally, a discontinuity in the equation of state of the scalar field,

\begin{equation}
    w_{\phi}=\frac{2}{3}\varepsilon-1.
\end{equation}

\begin{figure}
\centering
\includegraphics[]{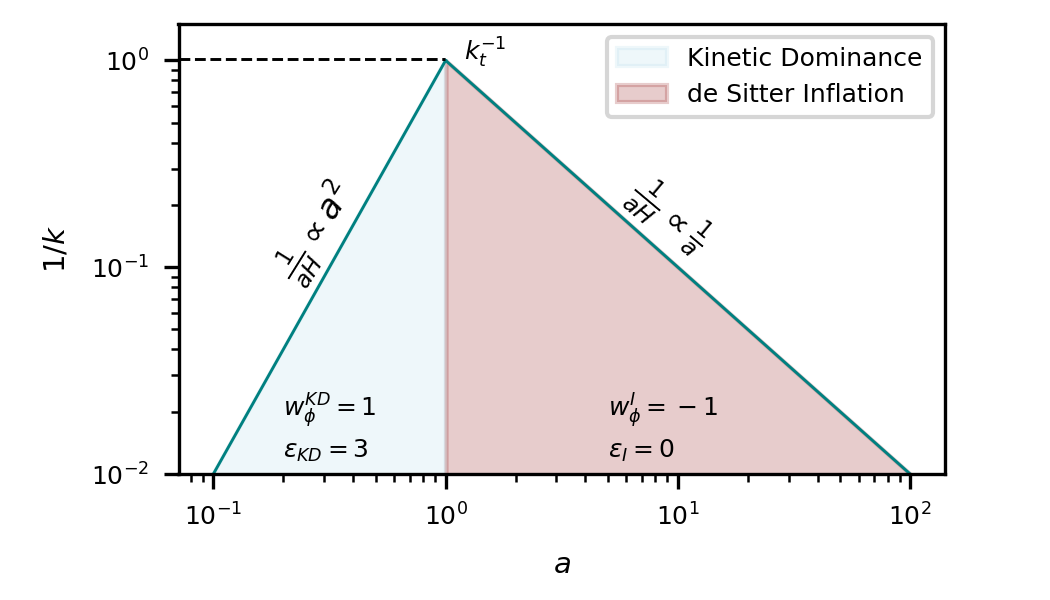}
\caption{Instantaneous transition in the comoving horizon between a period of kinetic dominance and de Sitter inflation as used in the Contaldi approximation. We have set $a=0$ at the Planck epoch; however, the convention that $a=1$ at the present epoch is not used but instead denotes the time of the instantaneous phase transition. Based on Fig. 3 in \cite{2021PhRvD.104f3532G}.
}
\end{figure}

\subsection{Kinetic dominance}
We refer to a slow-roll violating phase obeying \newline $\dot{\phi}^2\gg V(\phi)$ as kinetic dominance \cite{2003JCAP...07..002C, 2014PhRvD..89f3505H, 2016PhRvD..94b4041H, 2019PhRvD.100b3501H, 2019PhRvD.100b3502H}. The motivation for including this preinflationary phase follows from the original construction by \citet{2003JCAP...07..002C} so to provide an early Universe mechanism for suppression of the CMB power spectrum at low multipoles, $\ell$, as compared to that predicted by $\Lambda$CDM \cite{2014A&A...571A..22P,2014A&A...571A..15P}. This is, reduction in power at large scales is introduced via the primordial spectrum with a period of kinetic dominance. During such an epoch \cite{2003JCAP...07..002C},

\begin{equation}
\varepsilon_{\mathrm{KD}}=3.
\end{equation}

\noindent
This implies the following scaling of the comoving horizon:

\begin{equation}
    H_{\mathrm{KD}} \propto \frac{1}{a^3}.
\end{equation}

\noindent
Rearranging and changing to conformal time, one can obtain

\begin{equation}
   a_{\mathrm{KD}}(\tau)=\sqrt{1+2 k_{\mathrm{t}}\tau}.
\end{equation}

\noindent
Solving the MS equation approximation defined by (20) using the scale factor given above, results in the following primordial mode equation during kinetic dominance:

\begin{multline}
        v_{\mathrm{KD}}(\tau)=\sqrt{k\tau+\frac{k}{2kt}}\Bigg(A_k H_0^{(1)}\Big[k\tau+\frac{k}{2kt}\Big]\\
        +B_k H_0^{(2)}\Big[k\tau+\frac{k}{2kt}\Big]\Bigg),
\end{multline}

\noindent
where $A_k$ and $B_k$ are coefficients of integration representing the nonuniqueness of the primordial mode functions, and $H_0^{(1)}$, $H_0^{(2)}$ denote Hankel functions of the first and second kind.

\subsection{de Sitter inflation}
 de Sitter inflation is a regime defined by a constant Hubble parameter, $\dot{H}=0$ \cite{2009arXiv0907.5424B}, which immediately gives

\begin{equation}
   \varepsilon_{\mathrm{KD}}=0.
\end{equation}

\noindent
The comoving horizon then scales as

\begin{equation}
   \Big( \frac{1}{aH}\Big)_{\mathrm{I}}\propto \frac{1}{a}.
\end{equation}

\noindent
Rearranging and changing to conformal time, during de Sitter inflation the scale factor can be expressed as 

\begin{equation}
   a_{I}(\tau)=\frac{1}{1-  k_{\mathrm{t}} \tau}.
\end{equation}

\noindent
Again solving the MS equation approximation expressed by (20) using the scale factor defined by Eq. (28), the primordial mode equation during de Sitter inflation is

\begin{multline}
     v_{\mathrm{I}}(\tau)=C_k e^{-i(k\tau-\frac{k}{k_{\mathrm{t}}})}\Bigg(1-\frac{i}{k\tau-\frac{k}{k_{\mathrm{t}}}}\Bigg)\\
     +D_k e^{+i(k\tau-\frac{k}{k_{\mathrm{t}}})}\Bigg(1+\frac{i}{k\tau-\frac{k}{k_{\mathrm{t}}}}\Bigg),
\end{multline}

\noindent
where $C_k$ and $D_k$ are coefficients of integration.

\subsection{Analytic primordial power spectrum}

The primordial power spectrum is formed during the inflationary epoch when comoving curvature perturbations exit the comoving horizon and cease to evolve. The analytic primordial power spectrum in the Contaldi approximation can be derived from the dimensionless primordial power spectrum defined by Eq. (16) with use of the analytic functions for $v$ and $z$ during de Sitter inflation. The power spectrum becomes

\begin{equation}
    \mathcal{P_R}(k)=\lim_{\tau \rightarrow{}\frac{1}{k_{\mathrm{t}}}}\frac{k^3}{2\pi^2} \Big| \frac{v_I(\tau)}{z_I(\tau)}\Big|^2,
\end{equation}

\noindent
where the condition for the modes to be superhorizon, $k\ll aH$, is implemented by determining the value of $\tau$ at late times when all relevant modes are far outside the horizon \cite{2003JCAP...07..002C, 2021PhRvD.104f3532G}. From Eq. (28), for the scale factor in terms of conformal time during de Sitter inflation, the condition for $a\rightarrow{}\infty$  corresponds to $\tau \rightarrow{}\frac{1}{k_{\mathrm{t}}}$. Taking this limit in Eq. (30), the resulting analytic primordial power spectrum is expressed in terms of constants of integration for the primordial mode functions during de Sitter inflation which will be shown to depend on those during the kinetic dominance era. This is

\begin{equation}
     \mathcal{P_R}(k)=\frac{k_{\mathrm{t}}^2k}{4\pi^2\varepsilon_{\mathrm{I}}}|C_k-D_k|^2,
\end{equation}

\noindent
where $\varepsilon_{\mathrm{I}}$ is the first slow-roll parameter during inflation. $\varepsilon_{\mathrm{I}}$ should be set to zero to be consistent with the solution to the MS equation solved using approximations valid during a period of pure de Sitter inflation. Although, this would make the power spectrum divergent and in the case that $\varepsilon\ll1$, $H\approx \text{const.}$, Eqs. (27)-(28) describing the background for de Sitter inflation are still approximately true thus the perturbation mode functions defined by Eq. (29) may be used as a valid approximation \cite{2009arXiv0907.5424B}. In consequence, the Contaldi approximation will evaluate the power spectrum where choice of $\varepsilon_{\mathrm{I}}$ affects the amplitude of the power spectrum but not the scale dependence. The amplitude of the power spectrum can be absorbed into the parameter $A_s=\frac{k_{\mathrm{t}}^2}{4\pi^2\varepsilon_{\mathrm{I}}}$, with $k_{\mathrm{t}}=k_*$.

In order to obtain an expression for the analytic primordial power spectrum defined by Eq. (31), the functions $A_k$, $B_k$, $C_k$, and $D_k$ must be determined. The coefficients of integration for the kinetic dominance mode functions, $A_k$ and $B_k$, are solved by setting $v$ and $v'$ to quantum vacuum initial conditions such as those in \newline Table 1 \cite{2003JCAP...07..002C, 2021PhRvD.104f3532G}. Initial vacuum states are generally set far back in the inflationary epoch when all relevant modes are subhorizon.  If scales have initial conditions set at a time when they are not sufficiently deep within the comoving horizon, the choice of quantum vacuum may generate observationally distinguishable primordial power spectra \cite{2021PhRvD.104f3532G, 2005PhRvD..71j3512S}. Introducing a preinflationary phase of kinetic dominance such a consideration becomes important and lends itself to the decision of setting perturbation mode initial conditions at the time in which the comoving horizon is at a maximum using the equations for the kinetic dominance regime. A detailed treatment of observational consequences of choice of initial conditions is emphasized in the work of \citet{2021PhRvD.104f3532G}.

\begin{table}
\caption{Definitions of quantum vacuum perturbation mode initial conditions for Bunch-Davies vacuum (BD), Hamiltonian diagonalization (HD), renormalized stress energy tensor (RSET), and right-handed mode (RHM) \cite{2021PhRvD.104f3532G}.}
\begin{ruledtabular}
\begin{tabular}{ccccccc}
Initial condition &  &\\
\hline
\\
BD& $v_{k}=\frac{1}{\sqrt{2k}}$ & $v_k'=-ikv_k$ \\
HD& $v_{k}=\frac{1}{\sqrt{2k}}$ &$v_k'=-i\sqrt{k^2-\frac{z''}{z}}v_k$  \\
RSET& $v_{k}=\frac{1}{\sqrt{2k}}$ &$v_k'=(-ik+\frac{z'}{z})v_k $  \\
RHM&\multicolumn{2}{c}{$v_k(\tau)=\sqrt{\frac{\pi}{8k_{\mathrm{t}}}}\sqrt{1+2k_{\mathrm{t}}\tau}H_0^{(2)}\Big[k\tau+\frac{k}{2k_{\mathrm{t}}}\Big]$}
\end{tabular}
\end{ruledtabular}
\end{table}


To obtain the coefficients of integration of the primordial mode equations during the phase of de Sitter inflation, the scalar perturbations must be matched to the kinetic dominance era. These are fixed in the Contaldi approximation by imposing continuity of $v$ and $v'$ across the transition in regimes \cite{2003JCAP...07..002C}. The coefficients of integration $C_k$ and $D_k$ are then determined by equating the expression for $v$ and $v'$ in each era at the time of the transition. The absence of theoretical justification for propagating scalar primordial perturbations through a cosmological phase transition in this way initiates the need to derive physically acceptable cosmological matching conditions for the Contaldi approximation. 

Figure 2 shows the analytic primordial power spectra generated from the Contaldi approximations for BD, RHM, HD, and RSET vacuum initial conditions. The analytic expression for $C_k$ and $D_k$ are written out in Appendix C by Eqs. (C1)-(C8). A low $k$ cutoff exists around $\frac{k}{k_{\mathrm{t}}}\approx2$ for BD, RSET, and HD and at $\frac{k}{k_{\mathrm{t}}}\approx 1$ for RHM. Below the low $k$ cutoff the spectra experience power law distributions with BD and RSET $\propto k^2$, HD $\propto k^3$ and RHM $\propto k^3\mathrm{log}(k)^2$. In addition, there exists an intermediate region of damped oscillations before the spectrum plateaus at high $k$ values. The behavior at intermediate and high $k$ is all very similar for the initial conditions in consideration with the exception of RSET whose oscillations die down much more slowly before plateauing \cite{2021PhRvD.104f3532G}. The scale invariance (zero tilt) of the power spectrum is the result of the inflation phase being derived from approximations for a pure de Sitter regime. Note that it is intermediate values of $k$, which correspond to scales in the observable range \cite{2021PhRvD.104f3532G, 2014A&A...571A..22P, 2020A&A...641A..10P}.

\begin{figure}
\begin{center}
\includegraphics[]{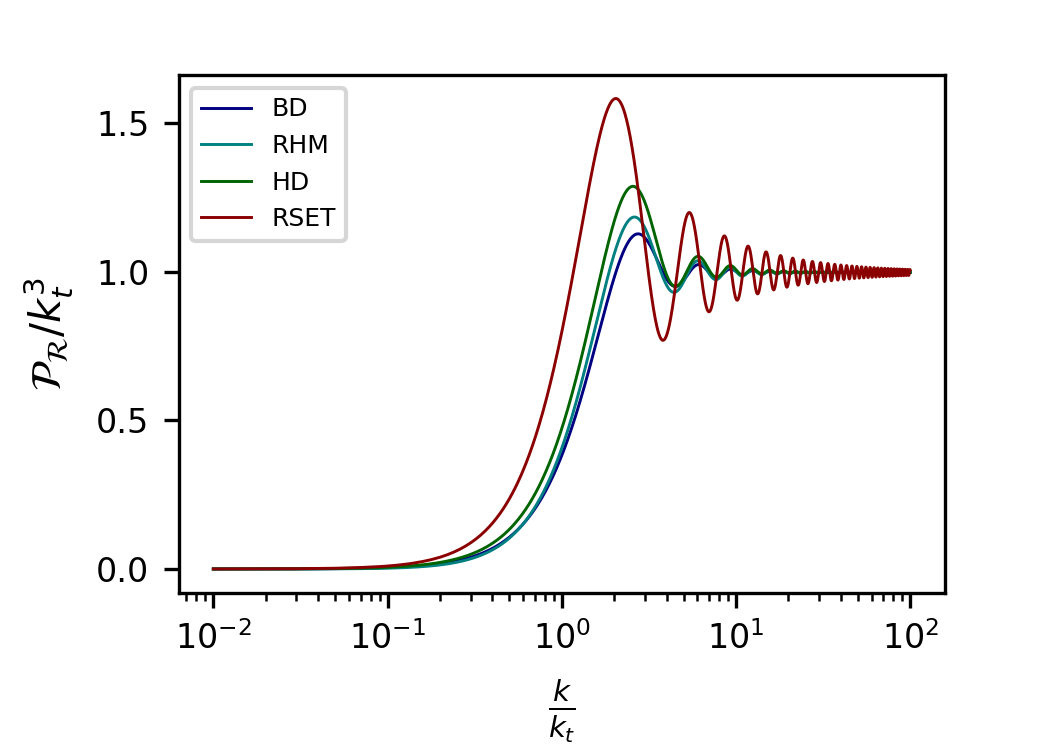}
\caption{Analytic primordial power spectra generated from the Contaldi approximation for BD, RHM, HD, and RSET vacuum initial conditions. $\varepsilon_{\mathrm{I}}$ is set to 0.0127 in order to normalize the power spectrum to 1. Based on Fig. 4 from \cite{2021PhRvD.104f3532G}.}
\end{center}
\end{figure}

\begin{figure}
\begin{center}
\includegraphics[]{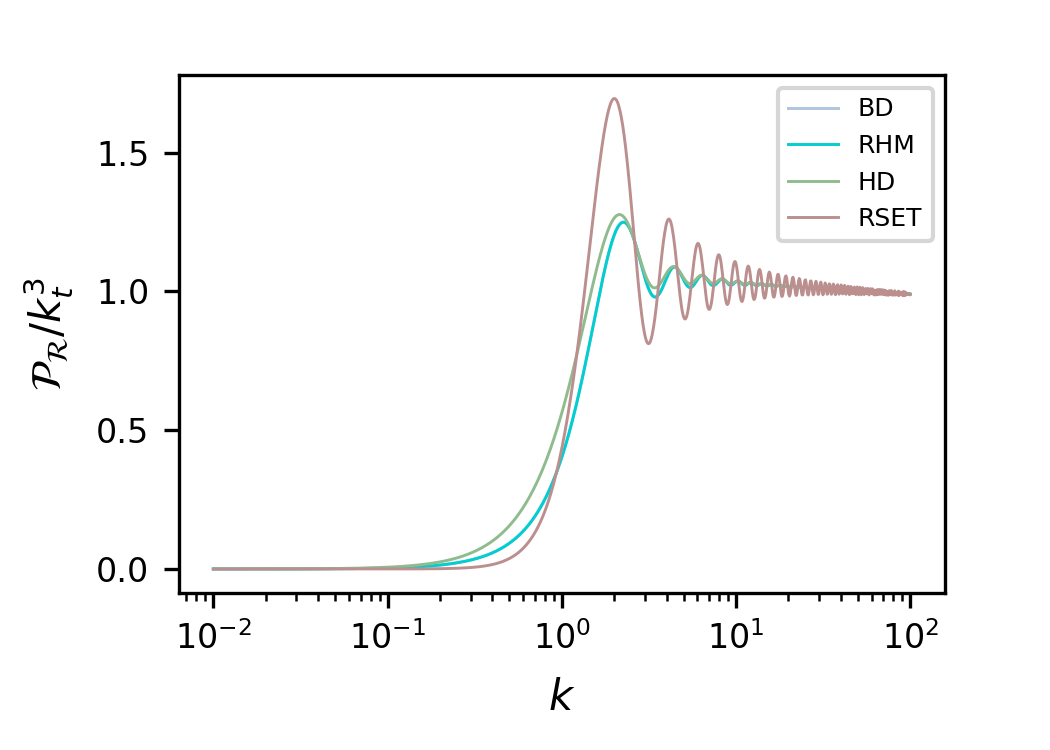}
\caption{Numerical primordial power spectrum for a \newline Starobinsky potential, $V(\phi)=\Lambda^4 \big(1-e^{-\sqrt{2/3}\phi}\big)^2$, with \newline $\Lambda^4 = 0.01$, where initial conditions for $\phi$, $\dot{\phi}$ have resulted in a preinflationary phase of kinetic domination. The spectra have been normalized to 1 for comparison with the spectra produced from the Contaldi approximation.}
\end{center}
\end{figure}

A full numerical evolution of the background equations and perturbation evolution equation allows for a comparison to the spectra produced above which does not require matching of the scalar perturbations across a jump in approximated background. Figure 3 shows primordial power spectra with background evolution arising from a Starobinsky inflationary potential where initial conditions for $\phi$, $\dot{\phi}$ have resulted in a preinflationary phase of kinetic domination. BD, RHM, HD, and RSET initial conditions for the perturbation equation are set at the maximum of the comoving horizon. The distinct behavior of each power spectra produced from applying the various initial conditions is comparable to those in Fig. 2, with the exception of RHM which looks much more similar to BD. The three regions of behavior of the power spectrum obtained by the Contaldi approximation are present in all spectra. These are, a power law at low $k$, damped region of oscillation in an intermediate regime and a plateauing at high $k$.

Figure 4 compares the primordial power spectrum produced from the Contaldi approximation and the numerical spectrum produced from the Starobinsky background for BD initial perturbation conditions. The spectra are very similar for small $k$ and both experience a low k cutoff at $k\approx 1$. The distinction emphasized in this plot is that the power spectrum produced from the Starobinsky inflationary potential with the given background initial conditions, $\phi$, and $\dot{\phi}$, results in a small tilt to the power spectrum which corresponds to a period of inflation that is not pure de Sitter with $H$ slowly varying. 

The similarity of the behavior of the analytic power spectra produced by the Contaldi approximation to the full numerical solutions would suggest that the joining of scalar perturbations across the instantaneous phase transition as done in the Contaldi model is the correct approach. We will nonetheless proceed with a precise analysis of acceptable general relativistic matching conditions to show this is not the case.

\begin{figure}
\centering
\includegraphics[]{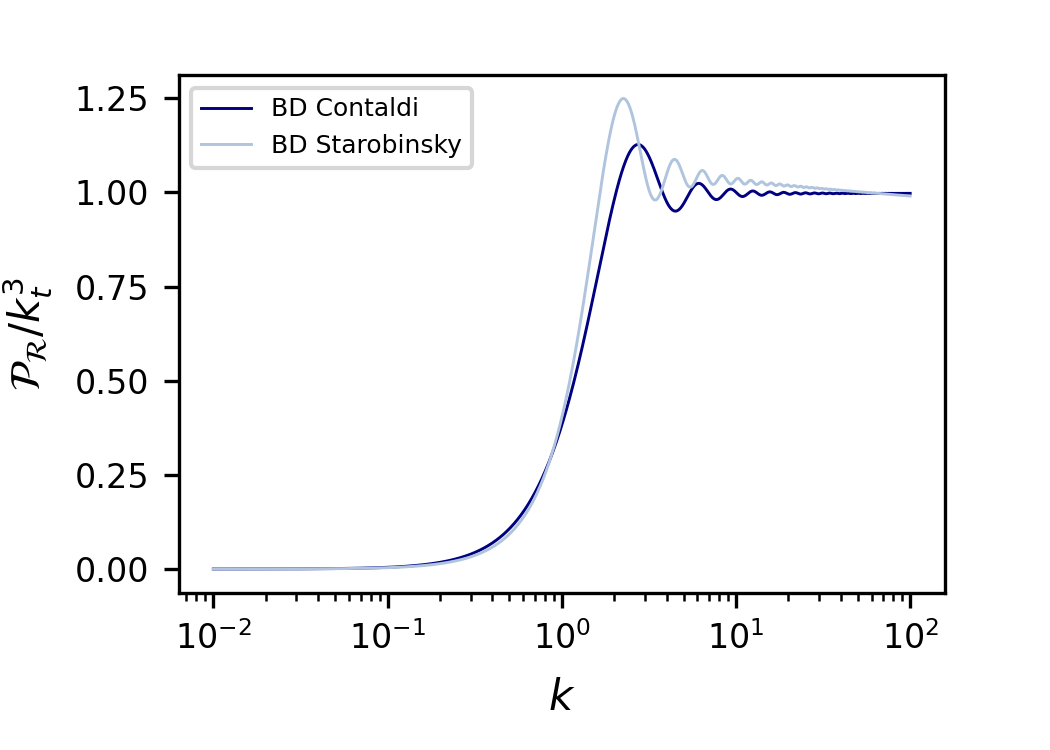}
\caption{Comparison of analytic primordial power spectrum from the Contaldi approximation and numerical primordial power spectrum from a Starobinsky potential for BD initial perturbation conditions. $k_{\mathrm{t}}$ is set to 1 in the Contaldi approximation.}
\end{figure}

\section{Cosmological matching conditions}

Despite the resemblance between the analytic and numerical primordial power spectra produced above from the Contaldi approximation and specification of an inflationary background, respectively, we wish to verify the use of physically consistent matching conditions for primordial perturbations which experience a jump in equation of state of the scalar field. We begin by introducing the junction conditions originally outlined by \citet{1967NCimB..48..463I}, which look at boundary surfaces and thin shells in general relativity to gain clarity regarding the appropriate treatment of surfaces of discontinuity. The proposed constraints allow the union of spacetime described by distinct metrics to smoothly join forming valid solutions to the Einstein field equations. The Israel junction conditions may be summarized as continuity of the first and second fundamental forms across the hypersurface, assumed not to be null, in the absence of a surface stress-energy tensor. For a complete derivation of the Israel junction conditions, one may refer to Appendix A. 

\subsection{Contaldi matching conditions}

In \citet{2003JCAP...07..002C}, the coefficients for the primordial mode functions during the de Sitter inflation era are obtained by requiring continuity of $v$ and $v'$ across the phase transition. We will refer to these as Contaldi matching conditions and in our notation are as follows. 

Defining a spacelike hypersurface for the transition $\Sigma:\tau=0$, continuity of the Mukhanov variable, $v$, across the hypersurface is 
\begin{equation}
    [v]_{\pm}=0.
\end{equation}

\noindent
For continuity of the first derivative of the Mukhanov variable in terms of conformal time, $v'$, across the hypersurface, we have

\begin{equation}
    [v']_{\pm}=0.
\end{equation}

\noindent
It should also be noted that the scale factor and the Hubble parameter are matched at the transition in the model, giving the additional constraints,

\begin{equation}
    [a]_{\pm}=[H]_{\pm}=0.
\end{equation}

\noindent
We will start with the formation of general cosmological matching conditions from the Israel junction conditions and then interpret the Contaldi matching conditions in light of such conclusions.

\subsection{Perturbation matching conditions from the Mukhanov-Sasaki equation}
We begin by deriving matching conditions for scalar perturbations with a method for effectively implementing the Israel junction conditions as has been done in much of the literature concerning propagating primordial perturbations through a jump in equation of state of the scalar field \cite{2012JCAP...12..018N,2016PhRvD..93l3519A,2019JCAP...06..028B}. Demanding the equation of motion for the comoving curvature perturbations does not contain singularities at the transition, the Israel junction conditions are assumed to be satisfied and cosmological matching conditions for the scalar perturbations can be obtained. The first requirement is continuity of the curvature perturbation itself,

\begin{equation}
    \big[\mathcal{R}\big]{\pm}=0.
\end{equation}

\noindent
The MS equation in terms of conformal time derivatives and the comoving curvature perturbation is

\begin{equation}
    \mathcal{R}''+2\frac{z'}{z}\mathcal{R}'+k^2\mathcal{R}=0.
\end{equation}

\noindent
In Sturm–Liouville form, this becomes

\begin{equation}
   \frac{d}{d\tau} \Big(\mathcal{R}'z^2\Big)=-z^2k^2\mathcal{R}.
\end{equation}

\noindent
Integrating both sides of Eq. (37) around the transition at $\tau=0$, where $\delta$ is a small displacement, we have

$$\int_{-\delta}^{+\delta}d \Big(\mathcal{R}'z^2\Big)=\int_{-\delta}^{+\delta}-z^2k^2\mathcal{R} d\tau.$$

\noindent
Recalling $z$ as defined in Eq. (19), 

\begin{equation}
[\mathcal{R}'z^2]_{\pm}=-2\int_{\delta}^{+\delta}a^2\varepsilon k^2\mathcal{R} d\tau.
\end{equation}

\noindent
The following change of variables can be made using the definition of the first-slow roll parameter,

\begin{equation}
    \varepsilon d\tau=-\frac{1}{a}d\Big(\frac{1}{H}\Big).
\end{equation}

\noindent
This substitution applied to Eq. (38) eliminates $\varepsilon$, which is the single parameter that jumps across the transition, and we arrive at the second cosmological matching condition, 

\begin{equation}
[\mathcal{R}'z^2]_{\pm}=0.
\end{equation}

\noindent
The two linearly independent matching conditions derived from this integral formulation are then

\begin{subequations}
    \begin{align}
      [\mathcal{R}]_{\pm}=0, \\
     [\mathcal{R}'z^2]_{\pm}=0.
    \end{align}
  \end{subequations}

\noindent
 It should be emphasized that matching conditions \newline (41a)-(41b) do not directly correspond to the first and second junction conditions respectively but are required to fulfil the conditions of continuity of the induced metric and extrinsic curvature, that which is not made clear in previous literature \cite{2012JCAP...12..018N,2016PhRvD..93l3519A,2019JCAP...06..028B}. In addition, a more careful investigation in the next section will show that assuming continuity of the comoving curvature perturbation given by condition (41a) amounts to making a choice for the definition of the hypersurface at the phase transition, which should not be held as trivial.

\subsection{Perturbation matching conditions defining a hypersurface at the transition}

We now implement the Israel junction conditions to obtain cosmological matching conditions for scalar primordial perturbations by explicitly defining a spacelike hypersurface at the transition and determining the functions which must be smooth in order for continuity of the induced metric and extrinsic curvature as demanded by the relevant constraints. We begin by making use of the work of \citet{1995PhRvD..52.5549D}, who sketch a procedure by which to derive cosmological matching conditions on a generic hypersurface . 

The general perturbed FRW metric in conformal time is

\begin{multline}
    ds^2=a(\tau)^2(1+2\Phi)d\tau^2-2B_idx^id\tau\\
    -[(1-2\Psi)\delta_{ij}+2E_{ij}+h_{ij}]dx^idx^j.
\end{multline}

\noindent
Suppose the stress-energy tensor which governs the evolution of Eq. (42) undergoes a finite discontinuity at a spacelike hypersurface $\Sigma: \varphi (\tau,x^i)=\bar{\varphi}(\tau)+\delta \varphi(\tau,x^i) =\text{const.}$, where $\varphi(\tau,x^i)$ is an arbitrary four-scalar with a homogeneous part, $\bar{\varphi}$ and a small inhomogeneous part, $\delta \varphi$. Under the coordinate transformation,

\begin{subequations}
    \begin{align}
      \tau \to \tilde{\tau}=\tau+\alpha, \\
      \tilde{x}^i=x^i+\delta^{ij}\beta_{,j}.
    \end{align}
  \end{subequations}

\noindent
The perturbation $\delta \varphi $ transforms as 

\begin{equation}
   \delta \varphi  \to \tilde{\delta \varphi }=\delta \varphi  -\bar{\varphi }'\alpha.
\end{equation}

\noindent
Going into the tilde coordinate system, $\tilde{\tau}=\text{const.}$ where $\tilde{\delta \varphi }=0$,
\begin{equation}
    \alpha=\frac{\delta \varphi }{\bar{\varphi }'}.
\end{equation}
\noindent
Immediately following from this is that the scale factor, $a$, and its first time derivative must be continuous across the hypersurface. From the first Israel junction condition, continuity of the metric defined by Eq. (42) implies the following two conditions in the tilde coordinate system:

\begin{equation}
    [\tilde{\Psi}]_{\pm}=0,
\end{equation}

\begin{equation}
    [\tilde{E}]_{\pm}=0.
\end{equation}

\noindent
From the second Israel junction conditions, continuity of the extrinsic curvature reads 
\begin{equation}
    [\tilde{\delta K^i_j}]_{\pm}=-\frac{1}{a}[\delta^i_j(\mathcal{H}\tilde{\Phi}+\tilde{\Psi}')+(\tilde{B}-\tilde{E}')^{,i}_{,j}]_{\pm}=0,
\end{equation}

\noindent
where the conformal Hubble parameter is $\mathcal{H}=\frac{a'}{a}$. 

Moving back into the original coordinate system gives matching conditions on $\Sigma: \bar{\varphi }+\delta \varphi =\mathrm{const.}$ in an arbitrary coordinate system,

\begin{subequations}
    \begin{align}
      [\Psi+\mathcal{H}\frac{\delta \varphi }{\bar{\varphi }'}]_{\pm}=0, \\
       [B-E'+\frac{\delta \varphi }{\bar{\varphi }'}]_{\pm}=0,   \\
       [\mathcal{H}\Phi +\Psi '+(\mathcal{H}'-\mathcal{H}^2)\frac{\delta \varphi }{\bar{\varphi }'}]_{\pm}=0.
    \end{align}
  \end{subequations}

\noindent
In the absence of anisotropic stress, the $ij$ Einstein equations give $\Phi=\Psi$. The following analysis will be done in the Newtonian/longitudinal gauge $(E=B=0)$ where the linearly independent conditions (49a)-(49c) for a hypersurface defined by an arbitrary scalar become

\begin{subequations}
    \begin{align}
      [\Psi]_{\pm}=0, \\
       [\frac{\delta \varphi }{\bar{\varphi }'}]_{\pm}=0,   \\
       [\mathcal{H}\Phi +\Psi '+(\mathcal{H}'-\mathcal{H}^2)\frac{\delta \varphi }{\bar{\varphi }'}]_{\pm}=0. 
    \end{align}
  \end{subequations}

Recovering cosmological matching conditions that can be applied to the joining of scalar primordial perturbations in the Contaldi approximation requires specification of the scalar, $\varphi$, defining the hypersurface at the transition between kinetic dominance and inflation. We now will consider the joining conditions emerging from two choices of $\varphi$.

\subsubsection{Hypersurface of constant energy density}

A hypersurface defining the cosmological phase transition in which the energy density, $\rho$, is constant expressed as $\Sigma:\bar{\rho}+\delta \rho=\text{const.}$ has been motivated in previous literature \cite{1995PhRvD..52.5549D, 1998PhRvD..57.3302M,2009PhRvD..80l3526A}. Equations (50a)-(50c) become

\begin{subequations}
    \begin{align}
      [\Psi]_{\pm}=0, \\
       [\frac{\delta \rho}{\bar{\rho}'}]_{\pm}=0,   \\
       [\mathcal{H}\Phi +\Psi '+(\mathcal{H}'-\mathcal{H}^2)\frac{\delta \rho}{\bar{\rho}'}]_{\pm}=0.
    \end{align}
  \end{subequations}

\noindent
Working in the Newtonian gauge, Eq. (51a) may be written as 

\begin{equation}
   [\Phi]_{\pm}=0.
\end{equation}

\noindent
 Equation (51b) can be rewritten so that $\bar{\rho}'$ and $\delta \rho$ are in terms of $a$ and $\mathcal{H}$. Using the $00$ linearized Einstein equations in the Newtonian gauge, 

\begin{equation}
    \bar{\rho}'=\frac{6\mathcal{H}}{ka^2}(\mathcal{H}'-\mathcal{H}^2),
\end{equation}

\begin{equation}
    \delta \rho=\frac{6}{ka^2}\Big[\frac{1}{3}\Delta \Phi -\mathcal{H}(\mathcal{H}\Phi+\Phi')\Big].
\end{equation}

\noindent
Equation (51b) is then

\begin{equation}
   [\frac{\delta \rho}{\bar{\rho}'}]_{\pm}=\Big[\frac{\mathcal{H}}{\mathcal{H}^2-\mathcal{H}'}(\Phi'+\mathcal{H}\Phi)+\frac{1}{3}\frac{\Delta \Phi}{(\mathcal{H}'-\mathcal{H}^2)}\Big]_{\pm}=0.
\end{equation}

\noindent
One can additionally obtain Eq. (56a) for cosmological perturbations in a universe dominated by a scalar field and the background Eq. (56b) from the Friedmann equations written in conformal time \cite{1992PhR...215..203M}, 

\begin{subequations}
    \begin{align}
      \Phi'+\mathcal{H}\Phi=\frac{1}{2} \delta \phi \bar{\phi}', \\
       \mathcal{H}'-\mathcal{H}^2=-\frac{1}{2} \bar{\phi}'^2 .
    \end{align}
\end{subequations}

\noindent
Redefining the comoving curvature perturbation from Eq. (9) using relations (56a)-(56b) gives

\begin{equation}
    \mathcal{R}=\Phi+\frac{\mathcal{H}}{\mathcal{H}'-\mathcal{H}^2}(\Phi'+\mathcal{H}\Phi).
\end{equation}

\noindent
From condition (55) and $\mathcal{R}$ as defined above, 

\begin{equation}
   [\frac{\delta \rho}{\bar{\rho}'}]_{\pm}=\Big[-\mathcal{R}+\Phi+\frac{1}{3}\frac{\Delta \Phi}{(\mathcal{H}'-\mathcal{H}^2)}\Big]_{\pm}=0.
\end{equation}

\noindent
Equation (51c) is then redundant and the linearly independent matching conditions for a hypersurface defined by constant $\rho$ are  

\begin{subequations}
    \begin{align}
    \Big[\mathcal{R}-\frac{1}{3}\frac{\Delta \Phi}{(\mathcal{H}'-
    \mathcal{H}^2)}\Big]_{\pm}=0,
    \\
 [\Phi]_{\pm}=0.
  \end{align}
\end{subequations}

\noindent
 Writing conditions (59a)-(59b) in terms of $\mathcal{R}$ and $z$, the following expression derived in  Appendix B is required:

\begin{equation}
    \Delta \Phi=\frac{1}{2}\frac{\bar{\phi'}^2}{\mathcal{H}}\mathcal{R}'.
\end{equation}

\noindent
 Using Eq. (56b) gives

\begin{equation}
    \frac{1}{3}\frac{\Delta \Phi}{(\mathcal{H}'-\mathcal{H}^2)}= -\frac{1}{3\mathcal{H}}\mathcal{R}'.
\end{equation}

\noindent
Additionally in Appendix B, the following relation is derived:

\begin{equation}
    \Phi=\frac{z^2H}{2ak^2}\mathcal{R}'.
\end{equation}

\noindent
With $k$, $H$, and $a$ matched across the transition, the cosmological joining conditions defined by (59a)-(59b) become

\begin{subequations}
    \begin{align}
    \Big[\mathcal{R}+\frac{1}{3\mathcal{H}}\mathcal{R}'\Big]_{\pm}=0,
    \\
 [z^2 \mathcal{R}']_{\pm}=0 . 
  \end{align}
\end{subequations}

Comparing with the matching conditions arrived at though the integral formulation in the previous section, condition (63a) differs from condition (41a) except in the long wavelength limit where the second term of Eq. (63a) may be ignored as $\mathcal{R}'$ is conserved. Condition (63a) is also equivalent to requiring the uniform-density curvature perturbation, $\zeta$, be continuous across the transition, where $\zeta=-\Psi+\frac{H}{\dot{\bar{\rho}}}\delta\rho$ \cite{2009arXiv0907.5424B, 2009PhRvD..80l3526A}. This is physically consistent as the hypersurface in consideration is one of uniform energy density.

\subsubsection{Hypersurface of constant scalar field}

An alternative choice of scalar defining the hypersurface at the transition is taking a surface of constant scalar field value \cite{2012JCAP...12..012C, 2012JCAP...12..018N, 2016PhRvD..93l3519A,  2007JCAP...06..014C, 1998PhRvD..57.3302M}. Expressing the hypersurface at the transition as
$\Sigma:\bar{\phi}+\delta \phi=\text{const.}$, the matching conditions take the form, 

\begin{subequations}
    \begin{align}
      [\Psi]_{\pm}=0, \\
       [\frac{\delta \phi}{\bar{\phi}'}]_{\pm}=0,   \\
       [\mathcal{H}\Phi +\Psi '+(\mathcal{H}'-\mathcal{H}^2)\frac{\delta \phi}{\bar{\phi}'}]_{\pm}=0 .
    \end{align}
  \end{subequations}

\noindent
Equation (64a) may again be written as

\begin{equation}
   [\Phi]_{\pm}=0.
\end{equation}

\noindent
Noting the definition of $\mathcal{R}$ from Eq. (9), condition (64b) can conveniently be taken in linear combination with constraint (65) as

\begin{equation}
    [\mathcal{R}]_{\pm}=0.
\end{equation}

\noindent
Additionally, recalling Eqs.  (56a)-(56b) it is clear that Eq. (64c) is trivially satisfied. The resulting cosmological matching conditions for a hypersurface defined by constant $\phi$ are

\begin{subequations}
    \begin{align}
    [\mathcal{R}]_{\pm}=0,
    \\
 [\Phi]_{\pm}=0.  
  \end{align}
\end{subequations}

\noindent
In terms of $\mathcal{R}$ and $z$, these are 

\begin{subequations}
    \begin{align}
    [\mathcal{R}]_{\pm}=0,
    \\
 [z^2\mathcal{R}']_{\pm}=0.
  \end{align}
\end{subequations}

These conditions are the same as that constructed via the MS equation expressed by constraints (41a)-(41b).

\subsection{A hypersurface for Contaldi matching}

In further consideration of the Contaldi matching conditions, one may look to see if the conservation of $v$ and $v'$ across the phase transition can be assigned as cosmological matching conditions for some  choice of hypersurface. This will be done by working backwards from conditions (32)-(33) in the Newtonian gauge to determine $\frac{\delta \varphi }{\bar{\varphi }'}$ for the generic cosmological matching conditions defined in Eqs.  (50a)-(50c) with corresponding hypersurface $\Sigma: \bar{\varphi}+\delta \varphi =\mathrm{const.}$. Noting the definition of the Mukhanov variable, $v$, the Contaldi matching conditions in terms of $\mathcal{R}$ become 

\begin{subequations}
    \begin{align}
    [z\mathcal{R}]_{\pm}=0,
    \\
 [z\mathcal{R}'+z'\mathcal{R}]_{\pm}=0. 
  \end{align}
\end{subequations}

\noindent
Recalling Eq. (62) for $\Phi$ in terms of $\mathcal{R}'$, Eq. (69b) may be written as 

\begin{equation}
\Big[\frac{2ak^2\Phi H}{z}+\frac{z'}{z}z\mathcal{R}\Big]_{\pm}=0. 
\end{equation}

\noindent
 Taking Eq. (69a) in linear combination with Eq. (70) with $\frac{z'}{z}$, $a$, $H$, and $k$ conserved across the hypersurface in the Contaldi approximation, the Contaldi matching conditions become

\begin{subequations}
    \begin{align}
    [z\mathcal{R}]_{\pm}=0,
    \\
 \Big[\frac{\Phi}{z}\Big]_{\pm}=0. 
  \end{align}
\end{subequations}

\noindent
The cosmological matching conditions for any choice of hypersurface requires $\Phi$ be matched across the transition as  stated by condition (50a). $[\Phi]_{\pm}=0$ can only be trivially satisfied for the matching of $v$ and $v'$ as there does not exist a condition for $z$ to independently be conserved across the transition. It is then that Contaldi matching does not correspond to a choice of hypersurface with cosmological matching conditions as it fails to satisfy the requirements of the Israel junction conditions. Importantly, this illustrates that the Contaldi matching conditions are not physically acceptable for reason that they do not account for the jump in $z$ from one regime to another, which is the result of a jump in the first-slow roll parameter, $\varepsilon$, or equally the scalar field equation of state, $w_{\phi}$, from kinetic dominance to de Sitter inflation. 

\subsection{On the choice of hypersurface}

This analysis has carefully considered the quantities which must not jump across a spacelike hypersurface defining a cosmological phase transition in order to ensure continuity of the first and second fundamental forms as demanded by the Israel junction conditions. By choosing different physical parameters to define the hypersurface of discontinuity the quantities that must be continuous across the transition differ on subhorizon scales. The matching conditions all reduce to Contaldi matching in the case that there is no jump in scalar field equation of state. This is clear in that $z$ becomes a conserved quantity across the transition; however, we are concerned with a cosmological scenario, which includes a phase transition defined by a jump in the equation of state of the scalar field and so the choice of hypersurface for the transition, which emits different matching conditions becomes crucial to constructing an accurate primordial power spectrum.

Justification for the choice of scalar defining the hypersurface of discontinuity is present in previous literature investigating the propagation of scalar perturbations through phase transitions \cite{2012JCAP...12..012C,2012JCAP...12..018N,2016PhRvD..93l3519A,2007JCAP...06..014C,1998PhRvD..57.3302M}. It is conveyed in \cite{1995PhRvD..52.5549D, 1998PhRvD..57.3302M,2009PhRvD..80l3526A}, that if the scalar field is an adiabatic perfect fluid, a jump in equation of state implies a jump in pressure and from the Friedmann equations the energy density remains constant. This lends itself to the choice that the hypersurface of discontinuity be $\Sigma: \bar{\rho}+\delta \rho=\text{const.}$ From \cite{2012JCAP...12..012C, 2016PhRvD..93l3519A, 2007JCAP...06..014C}, it is stressed that if a transition in equation of state is triggered by a local physical quantity, the hypersurface must be a function of of $\phi$, suggesting $\Sigma: \bar{\phi}+\delta \phi=\text{const.}$ Although both choice of scalars defining the hypersurface look to be allowable, there remains no theoretical motivation for a canonical definition for the hypersurface of transition. 

\section{Primordial power spectrum with cosmological matching conditions}

We now consider the behavior of the primordial power spectra produced by applying the two sets of cosmological matching conditions derived in the previous section to the Contaldi approximation. 

Figure 5 shows the primordial power spectrum resulting from the Contaldi approximation using cosmological conditions (63a)-(63b) arrived at by applying Israel junction conditions to a hypersurface of constant energy density, $\rho$, defining the transition between kinetic dominance and inflation. The coefficients of integration $C_{k}$ and $D_k$ are written out in Appendix C in Eqs.  (C9)-(C16). The behavior of the power spectrum is clear through looking at these expressions. The amplitude of the power spectrum is modified, where there exists a scaling of $\varepsilon_{\mathrm{I}}^{-2}$ as compared to $\varepsilon_{\mathrm{I}}^{-1}$ which is present in Contaldi matching. This alters the normalization of the power spectrum. Enhancement of oscillations which are no longer damped at high $k$ correspond to  $v_{\mathrm{I}}$ and $v'_{\mathrm{I}}$ being in phase. Moreover, the power spectrum is no longer scale independent as leading order in $k$ has become $\sqrt{k}$ rather than $\frac{1}{\sqrt{k}}$ as in Contaldi matching. This gives a $k^2$ dependence of the primordial power spectrum recalling Eq. (31).

\begin{figure}
\centering
\includegraphics[]{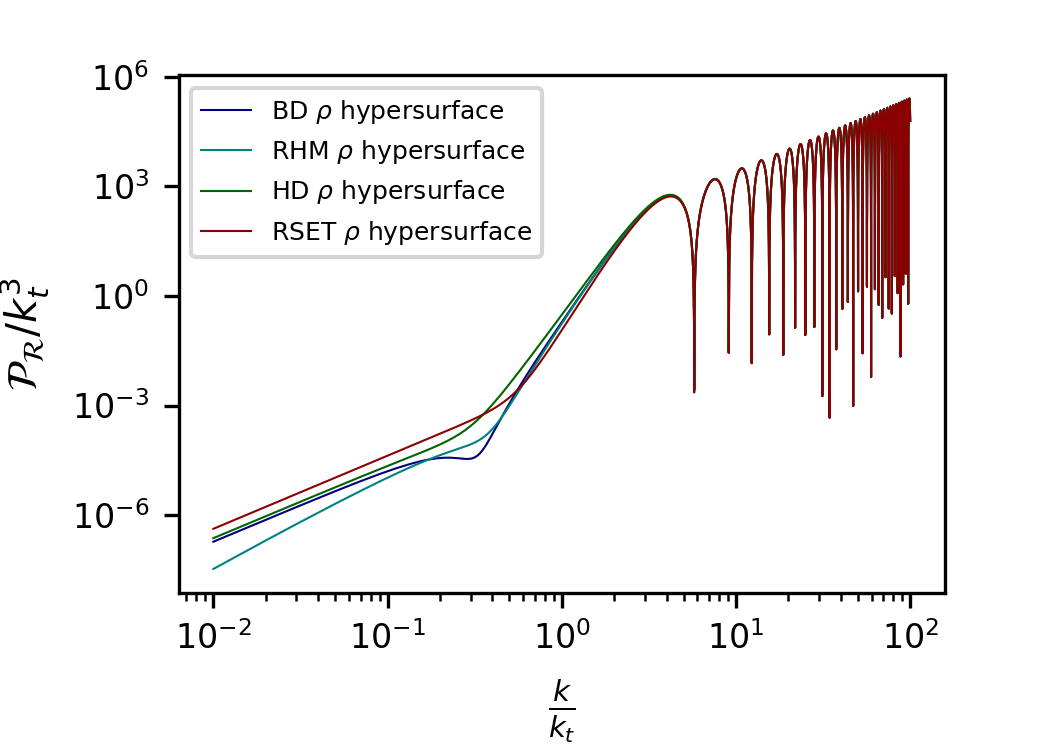}
\caption{Analytic primordial power spectra produced from the Contaldi approximation using cosmological matching conditions for a hypersurface defined by $\Sigma:\bar{\rho}+\delta \rho=\text{const.}$ BD, RHM, HD, and RSET initial conditions are shown. $\varepsilon_{\mathrm{I}}$ has been set to 0.0127 for comparison with Contaldi matching}
\end{figure}

Figure 6 gives the primordial power spectrum generated by the Contaldi approximation using cosmological conditions (68a)-(68b) resulting from Israel junction conditions applied to a hypersurface of constant scalar field value, $\phi$, defining the transition which coincide with those arrived at through considering singular terms in the MS equation. The coefficients of integration $C_k$ and $D_k$ are written out in Appendix C in Eqs.  (C17)-(C24).  As with the choice of a constant energy density hypersurface, the amplitude of the power spectrum changes due to a $\varepsilon_{\mathrm{I}}^{-2}$ dependence.
This agrees with conclusions from \citet{2012JCAP...12..012C}. Oscillations no longer plateau at high $k$ due to $v_{\mathrm{I}}$ and $v'_{\mathrm{I}}$ being in phase. Importantly, scale invariance is retained as leading order in $k$ remains $\frac{1}{\sqrt{k}}$.

Although the behavior of the power spectra produced from applying the cosmological matching conditions to the Contaldi approximation falls directly from the joined coefficients, $C_k$ and $D_k$, unphysical features in the spectra that result from applying such physically motivated conditions suggests a closer look should be taken at the impact of instantaneous transitions on the primordial power spectrum. We do this by presenting an alternative model to the Contaldi approximation which generates primordial power spectra from background evolution permitting arbitrary sharp cosmological phase transitions.

\section{A smooth semianalytic model for the primordial power spectrum}

\begin{figure}
\centering
\includegraphics[]{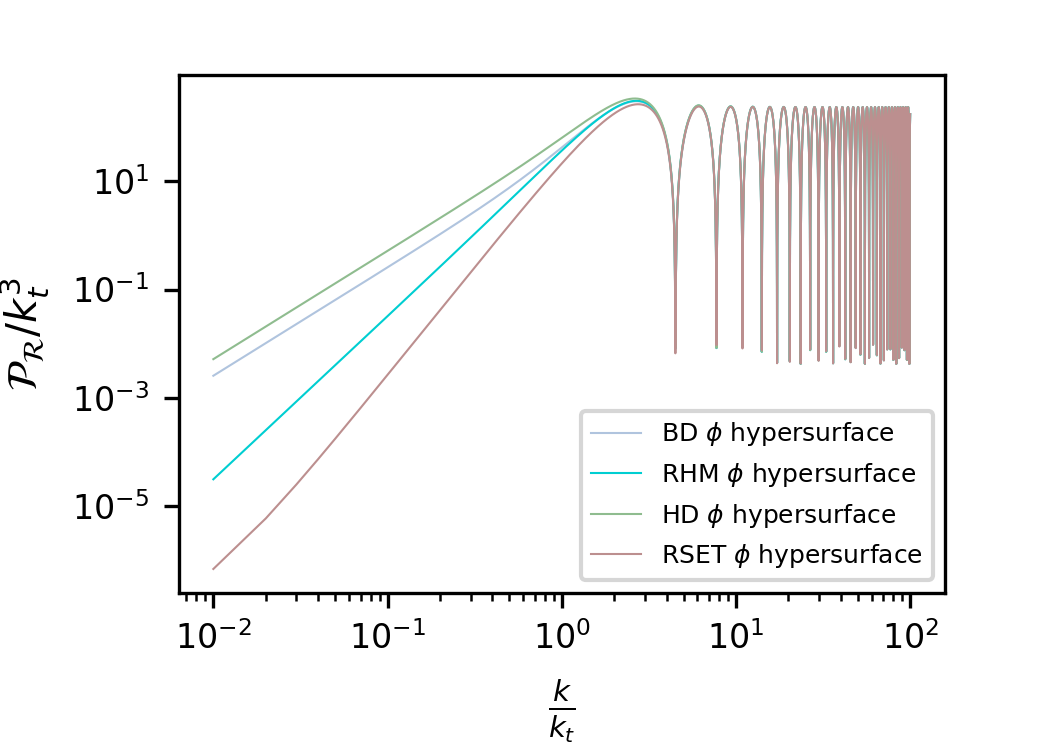}
\caption{Analytic primordial power spectra produced from the Contaldi approximation using cosmological matching conditions for a hypersurface defined by $\Sigma:\bar{\phi}+\delta \phi =\text{const.}$ BD, RHM, HD, and RSET initial conditions are shown. $\varepsilon_{\mathrm{I}}$ has been set to 0.0127 for comparison with Contaldi matching.}
\end{figure}

The following section details a novel semianalytic method for computing the primordial power spectrum. This is done by smoothly joining the approximations to the comoving horizon for a phase of kinetic dominance and inflation. With analytic solutions to the background Universe one may express the MS equation analytically giving an expression which can be solved numerically. The motivations for this approach are threefold; the produced primordial power spectrum remains independent of a choice of the inflationary potential, it does not require matching conditions for the scalar perturbations, and finally, control is gained over the duration of the cosmological phase transition. The catalyst for this model is both in an alternative to the Contaldi approximation and the ability to produce a power spectrum from an arbitrarily sudden cosmological transition which will prove useful for comparing to spectra produced from applying cosmological matching conditions to an instantaneous transition in the Contaldi approximation. 

\subsection{Constructing a solution with pure de Sitter inflation}

The  MS equation in terms of analytic functions of $H(N)$ and $z(N)$ and derivatives with respect to number of $e$-folds, $N=\mathrm{log}a$, is

\begin{equation}
    \mathcal{R}''+\Big(1+\frac{H'(N)}{H(N)}+2\frac{z'(N)}{z(N)}\Big)\mathcal{R}'+\Big(\frac{k}{a(N)H(N)}\Big)^2 \mathcal{R}=0.
\end{equation}

\noindent
 The expressions for $H(N)$, $\frac{H'(N)}{H(N)}$, $z(N)$ and $\frac{z'(N)}{z(N)}$ may be obtained by setting an analytic equation for the comoving horizon. This is done via the following procedure:

$$
 \text{Set } \frac{1}{aH}\xrightarrow{\mathrm{solve}}H\xrightarrow{\mathrm{differentiate}}\frac{H'}{H}.$$

The comoving horizon during kinetic dominance scales as

$$\Big( \frac{1}{aH}\Big)_{\mathrm{KD}}\propto a^2,$$

\noindent
and that during de Sitter inflation is 

$$\Big( \frac{1}{aH}\Big)_{\mathrm{I}}\propto \frac{1}{a}.$$

\noindent
The smooth comoving horizon obtained by combining the scaling of the horizons in kinetic dominance and de Sitter inflation may be generalized to produce increasingly sharp transitions by introducing a parameter, $\alpha \in \mathbb{R}_{>0}$, giving a comoving horizon with the resulting functional form, 

\begin{equation}
    \frac{1}{aH(a)}= \frac{1}{(a^{\alpha}+\frac{1}{a^{2\alpha}})^{1/\alpha}}.
\end{equation}

\noindent
In terms of $e$-folds, the comoving horizon is defined as 

\begin{equation}
    \frac{1}{a(N)H(N)}=\Big(e^{-2\alpha N}+e^{\alpha N}\Big)^{-\frac{1}{\alpha}}.
\end{equation}

\noindent
$H(N)$ can be obtained by rearranging 

\begin{equation}
    H(N)= \Big(1+e^{-3\alpha N} \Big)^{\frac{1}{\alpha}}.
\end{equation}

\noindent
Differentiating with respect to $e$-folds gives

\begin{equation}
    \frac{H'(N)}{H(N)}=-\frac{3}{1+e^{3\alpha N}}.
\end{equation}

\noindent
The first slow-roll parameter defined in terms of derivatives with respect to $e$-folds is

\begin{equation}
    \varepsilon(N)=-\frac{H'(N)}{H(N)}.
\end{equation}

\noindent
Therefore, the analytic expression for the first-slow roll parameter with the background specified by Eq. (74) is 

\begin{equation}
    \varepsilon(N)=\frac{3}{1+e^{3\alpha N}}.
\end{equation}

Noting that Eq. (78) is divergent, resulting from the stage of pure de Sitter inflation, a primordial power spectrum cannot be constructed using this background. We proceed by implementing a non-de Sitter inflationary stage obtained through modifying the functional form of the first slow roll-parameter. 

\subsection{Constructing a solution with modified de Sitter inflation}

 An equation for $\varepsilon(N)$ which does not tend to zero during the period of inflation can be determined by using the following reverse procedure:

\begin{equation}
    \text{Modify } \varepsilon=-\frac{H'}{H}\xrightarrow{\mathrm{solve}}H\xrightarrow{}\frac{1}{aH},
\end{equation}

\noindent
giving a comoving horizon which smoothly connects an epoch of kinetic dominance to a period of modified de Sitter inflation. This so-called modified de sitter inflation era is characterized by the slow-roll parameters $\varepsilon_{\mathrm{I}}, |\eta(N)| \ll 1$, where $\eta$ is the second slow roll parameter. These conditions capture a slowly decreasing Hubble parameter.

We modify Eq. (78) to take the following form:

\begin{equation}
    \varepsilon(N)=\frac{3-\varepsilon_{\mathrm{I}}}{1+e^{3 \alpha N}}+\varepsilon_{\mathrm{I}},
\end{equation}

\noindent
 such that a nonzero first slow-roll parameter is attained at the end of the finite duration phase transition, $ \lim_{N\to\infty} \varepsilon(N)=\varepsilon_{\mathrm{I}}\neq 0$. The general form of the background equations can be solved starting from the following equation, which has derivatives in terms of $e$-folds,

\begin{equation}
    \frac{H'(N)}{H(N)}=\frac{-3+\varepsilon_{\mathrm{I}}}{1+e^{3 \alpha N}}-\varepsilon_{\mathrm{I}}.
\end{equation}

\noindent
Solving for $H(N)$, the relevant equations become

\begin{equation}
    H(N)=e^{-3N}(1+e^{3\alpha N})^{\frac{3-\varepsilon_{\mathrm{I}}}{3\alpha}},
\end{equation}

\begin{equation}
    \frac{1}{a(N)H(N)}=e^{2N}(1+e^{3\alpha N})^{\frac{-3+\varepsilon_{\mathrm{I}}}{3\alpha}},
\end{equation}

\begin{equation}
    z(N)=e^N\sqrt{\frac{6+2\varepsilon_{\mathrm{I}}e^{3 \alpha N}}{1+e^{3\alpha N}}},
\end{equation}

\begin{equation}
    \frac{z'(N)}{z(N)}=\frac{6+e^{3 \alpha N}\Big(3\alpha (-3+\varepsilon_{\mathrm{I}})+2(3+\varepsilon_{\mathrm{I}}+\varepsilon_{\mathrm{I}}e^{3 \alpha N})\Big)}{2(1+e^{3\alpha N})(3+\varepsilon_{\mathrm{I}}e^{3 \alpha N})}.
\end{equation}

\noindent
Equations (81), (83), and (85) comprise the analytic equations necessary to write Eq. (72) fully analytically; however, the functional forms of these expressions require the MS equation be solved numerically. Evolving perturbations until they are superhorizon, a numeric primordial power spectrum can be generated from the constructed background.

The second slow roll-parameter in terms of derivatives with respect to $e$-folds is

\begin{equation}
  \eta(N)=\varepsilon(N)-\frac{\varepsilon'(N)}{2\varepsilon(N)}.
\end{equation}

\noindent
After the transition from kinetic dominance, this model assumes a constant first-slow roll parameter, $\varepsilon_I$. That is $\frac{\varepsilon'(N)}{\varepsilon(N)}=0$, which demands the second slow-roll parameter evaluated at observationally relevant $k$ follow from the choice of $\varepsilon_{\mathrm{I}}$ and $\alpha$ where $\lim_{N\to\infty} \eta(N)=\varepsilon_{\mathrm{I}}$. Allowing for a time varying first slow-roll parameter during inflation gives an end to inflation and results in a model for producing the primordial power spectrum where the first and second slow-roll parameters at the pivot scale can be set. A sketch of the procedure by which to obtain such a model is present in Appendix D; however, the analytic background presented in this section is sufficient for our considerations which concentrate on attaining spectra from sudden finite transition to compare to those of instantaneous cosmological transition produced in the Contaldi approximation.

\subsection{Cosmological phase transition duration}

In the semianalytic model we have presented, the duration of the cosmological phase transition can be approximated by looking at Eq. (80) giving the analytic equation for the first slow-roll parameter and determining the number of $e$-folds it takes to change from $\varepsilon_{\mathrm{KD}}$ to $\varepsilon_{\mathrm{I}}$. We shall define the cosmological transition as when $\varepsilon(N)$ is further than $1\%$ away from the associated value of the first slow-roll parameter during the kinetic dominance and inflation epochs. This choice captures the difference in the length of the transition with a change in  $\varepsilon_{\mathrm{I}}$, which is not encompassed by simply requiring the slow-roll conditions are met. The end of the period of kinetic dominance corresponds to the start of the phase transition,

\begin{equation}
    \varepsilon(N)\le \varepsilon_{\mathrm{KD}}-0.01\varepsilon_{\mathrm{KD}},
\end{equation}

\noindent
and the beginning of the modified de Sitter inflation period is equally the end of the cosmological phase transition,

\begin{equation}
    \varepsilon(N)\le \varepsilon_{\mathrm{I}}+0.01\varepsilon_{\mathrm{I}}.
\end{equation}

\noindent
The cosmological phase transition then occurs when  \newline $\varepsilon_{\mathrm{I}}+0.01\varepsilon_{\mathrm{I}}<\varepsilon(N)<\varepsilon_{\mathrm{KD}}-0.01\varepsilon_{\mathrm{KD}}$.

\begin{figure}
\centering
\includegraphics[]{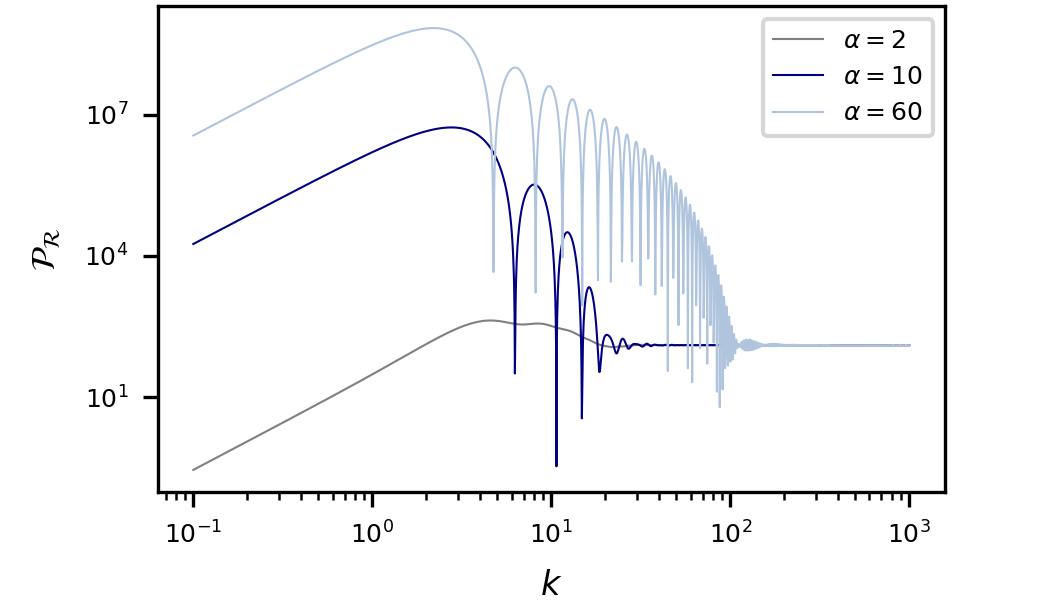}
\caption{Numerical primordial power spectra for increasingly sharp cosmological phase transitions generated by the semianalytic model presented in this section. $\varepsilon_{\mathrm{I}}=0.0001$ and $\alpha=2,10,60$ with BD initial conditions for perturbation modes set at the maximum of the analytic comoving horizon. The duration of the transitions are 3.25, 0.65, and 0.12 $e$-folds, respectively.}
\end{figure}

Figure 7 shows primordial power spectra produced from the model presented in this section with a change in the duration of the cosmological transition which is characterized by $\alpha$ in Eqs.  (81)-(85) for the background. It is evident that the length of the phase transition has a large effect on the resulting spectra.  An approximate duration of the cosmological transition for each background can be calculated by taking the difference between the end of the kinetic dominance period and the beginning of inflation defined by Eqs.  (87) and (88). The spectra are all identical at sufficiently large k; however, distinct behavior is particularly noticeable at intermediate $k$, where for a sufficiently fast transition there exists an enhancement of oscillations at some scales. Importantly, this intermediate region of $k$ corresponds to the observationally relevant scales. The change in behavior occurs for a greater range in scale, the shorter the transition duration. That is, oscillations begin to be enhanced at the same scale but are effected up to higher $k$ when the cosmological phase transition occurs over a shorter duration. Additionally, although we only show the primordial power spectrum with BD conditions in Fig. 7, the affects of choice of initial conditions for the perturbation modes is more pronounced at observationally relevant scales the shorter the duration of the transition.

Figure 8 compares the primordial power spectrum produced from an instantaneous transition in the Contaldi approximation using cosmological matching conditions for a transition hypersurface defined by constant $\phi$ and the power spectrum produced from a sufficiently fast finite duration phase transition. For a specified range of $k$, the power spectra differ by less than $1\%$. The scales at which this occur are those corresponding to enhanced scales in the sudden finite duration transition power spectrum. The conditions for which a power spectrum resulting from a finite duration transition looks like that produced from an instantaneous transition has been consider by \citet{2012JCAP...12..012C} and \citet{2016PhRvD..93l3519A} as follows.

\begin{figure}
\centering
\includegraphics[]{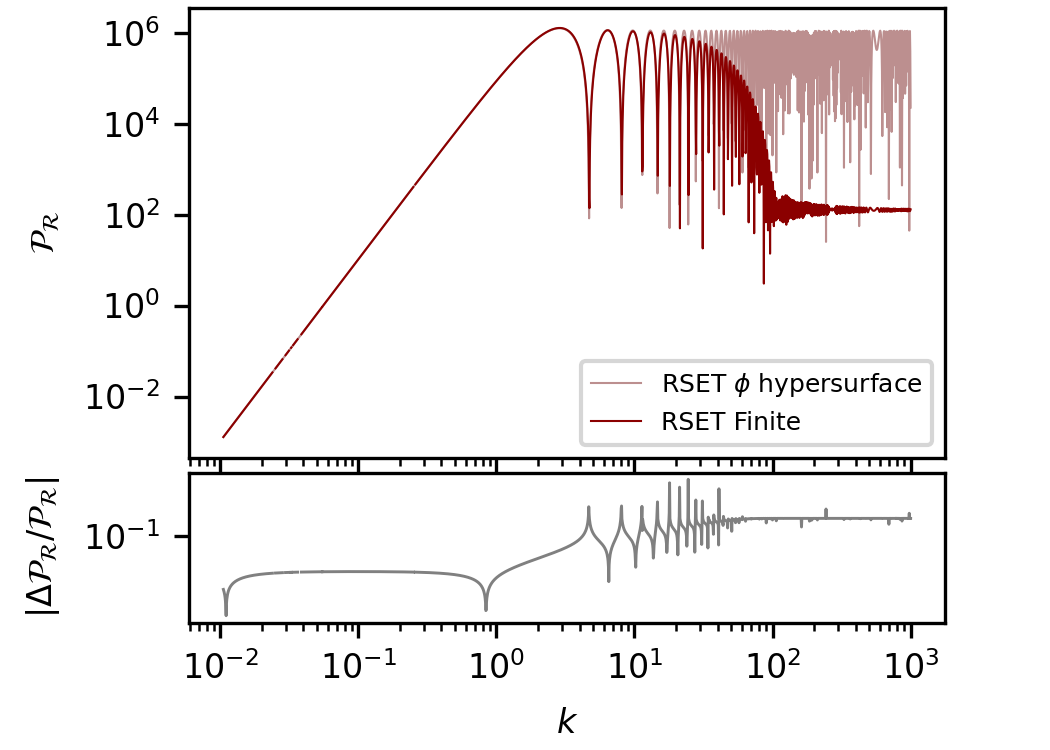}
\caption{The upper plot compares primordial power spectra produced from an analytic background with a sufficiently sudden duration phase transition defined by Eq. (83) taking $\alpha=60$ and that from cosmological matching conditions with hypersurface of constant $\phi$ applied to the instantaneous phase transition in the Contaldi approximation. $\varepsilon_{\mathrm{I}}$ is set to 0.0001 and $k_t$ is set to 0.98945 in the analytic power spectrum to align with the background of the numerical solution. RSET initial conditions have been used. The lower plot shows the percent error which remains small for scales obeying the condition expressed by (89).}
\end{figure}

A primordial power spectrum produced from a cosmological scenario which transitions between inflation and a slow-roll violating phase over a timescale, $T$, can be approximated by a primordial power spectrum produced from an instantaneous transition for scales obeying

\begin{equation}
    \frac{k}{a_{\mathrm{t}}}\ll \frac{1}{T},
\end{equation}

\noindent
where $a_t$ is the scale factor at the maximum of the analytic comoving horizon defined by Eq. (83). In the limit of an instantaneous transition, all modes will be enhanced by the phase transition. 

Primordial power spectra produced with no enhancement in oscillations at intermediate $k$ occur for transition lengths on the order of a single $e$-fold depending on choice of $\varepsilon_{\mathrm{I}}$; thus, if the cosmological phase transition is not sudden it does not imprint on the primordial power spectrum. This may explain why the power spectra produced by the Contaldi approximation using Contaldi matching in Fig. 2 looks similar to numerical power spectra computed with $\mathcal{O}(1)$ transitions specified by an inflationary potential in Fig. 3.  Specifically, it has been concluded that Contaldi matching does not join the primordial scalar perturbations in a way that takes into account the instantaneous jump in equation of state of the scalar field. The resulting primordial power spectrum that does not encode the effects of an instantaneous phase transition could reasonably be expected to exhibit similar behavior to a power spectrum produced by a transition which is too slow to enhance the power spectrum at any relevant scale, as quantified by Eq. (89). Cosmological phase transitions are thought to occur over the duration of several $e$-folds \cite{1998PhRvD..57.3302M}. The difference between primordial power spectrum produced from slow and sudden transitions suggests that producing a primordial power spectrum from a background which is described by an instantaneous transition should be done with caution if cosmological phase transitions are thought to happen over longer timescales. 

We then suggest that the procedure for producing the primordial power spectrum presented here  may be used as an alternative model to the Contaldi approximation which allows for greater control over both the scale dependence of the power spectrum through specification of $\varepsilon_{\mathrm{I}}$ and the duration of the cosmological phase transition controlled by $\alpha$. Although we have introduced this model as a potential independent method for computing the primordial power spectrum, the following section considers the implicit inflationary potential of the background model. 

\subsection{Scalar field potential reconstruction} 

The Hamilton-Jacobi formalism treats the Hubble parameter as the fundamental quantity changing with time. This approach allows for reconstruction of a scalar field potential, $V(\phi)$, for a specified $H(\phi)$. In terms of derivatives with respect to number of $e$-folds, this is

\begin{equation}
    V(N)=3H(N)^2+H'(N)H(N).
\end{equation}

\noindent
Specifying $H(N)$ for a cosmological evolution smoothly joining an era of kinetic dominance with inflation as denoted by Eq. (82), gives the reconstructed potential which in turn admits the equation $H(N)$ as an exact inflationary solution. A function for $\phi(N)$ can be used in order to write the potential as a function of the scalar field. Changing Eq. (18) into derivatives with respect to number of $e$-folds gives the following equation which may be solved to obtain $\phi(N)$,

\begin{equation}
    \phi(N)=\phi_0\pm\int \sqrt{-2\frac{H'(N)}{H(N)}}dN.
\end{equation}

Plotting the potential expressed by Eq. (90)  parametrically as a function of the solution to Eq. (91) gives the behavior of the associated $V(\phi)$ for a background specified by Eq. (83).

Figure 9 shows the reconstructed potential for this model with $\varepsilon_{\mathrm{I}}=0.0001$ and $\alpha=1$. Taking note of the region of the potential in which inflation occurs, the potential produced from these parameters is a convex ($V_{,\phi\phi}>0$), small field potential where the conditions for inflation are met in that the magnitude of the first and second derivatives of the potential with respect to $\phi$, $V_{,\phi}$ and $V_{,\phi \phi}$, are small \cite{2020A&A...641A..10P, 2009arXiv0907.5424B}. The convex form of this potential is characteristic for the associated background for $\varepsilon_{\mathrm{I}}\ll 1$ and $\alpha<3$, which corresponds to transitions of at least one $e$-fold in duration. The form of this potential is not ruled out observationally as in the case of usual convex large field inflationary potentials through high values of the tensor-to-scalar ratio, $r$ \cite{2009arXiv0907.5424B, 2020A&A...641A..10P}. 

\begin{figure}
\centering
\includegraphics[]{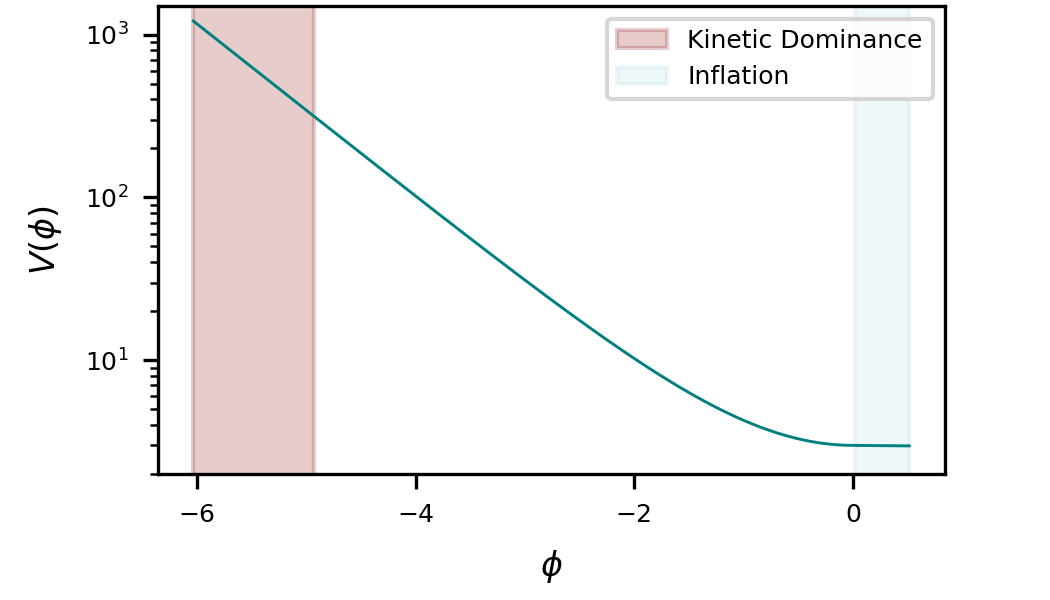}
\caption{Characteristic convex inflationary potential from parametric reconstruction for a smooth comoving horizon defined by Eq. (83) with a slow transition corresponding to $\varepsilon_{\mathrm{I}}\ll 1$ and $\alpha<3$. This example has
$\alpha=1$ and $\varepsilon_{\mathrm{I}}=0.0001$. The regions of the potential corresponding to kinetic dominance and inflation specified using Eqs.  (87) and (88).}
\end{figure}

Figure 10 shows the reconstructed potential for this model with $\varepsilon_{\mathrm{I}}=0.0001$ and $\alpha=10$. The corresponding primordial power spectrum is seen in Fig. 7. This example again results in a small field inflationary potential; however, the form of the potential is close to that of a step function which is not supported by current observation. Moreover, it is unphysical to require the scalar field be pushed up the potential towards inflation. Potentials of this form are generic for $\varepsilon_{\mathrm{I}}\ll 1$ and $\alpha>3$, where the steplike potential is required to achieve a transition on the timescale of fractions of an $e$-fold. Note that the Contaldi approximation has an implicit inflationary potential which is a Heaviside step function. 

Through the Hamilton-Jacobi formalism, one can see that although the semianalytic model we have presented in this section does not demand a choice for the inflationary potential to emit the desired background evolution, the generic construction has associated an implicit potential. This allows for phenomenological analyses in which an observationally constrained primordial power spectrum admits an acceptable functional form for the inflationary potential.

\section{Conclusion}
We have considered analytic and numeric procedures for generating the power spectrum of primordial scalar perturbations in the case of a background Universe which undergoes a jump in equation of state of the scalar field. Although there exits interest in a closed universe \cite{2021PhRvD.103d1301H},  we consider the simplest case of a flat universe for the purpose of isolating our conclusions wherein further analysis can be used to extended the results to the case of a curved universe. The Contaldi approximation provides an inflationary potential independent method for producing the primordial power spectrum by implementing an instantaneous transition between a phase of kinetic dominance and de Sitter inflation where approximate primordial mode equations exist. The aim of this analysis is the application of Israel junction conditions to determine the physically acceptable way in which to propagate primordial scalar perturbations across cosmological phase transitions allowing for clarification of previous work. The resulting joining conditions are seen to require specification of the scalar defining the spacelike hypersurface at the transition. Cosmological matching conditions corresponding to hypersurfaces of constant scalar field value and energy density were theoretically motivated. Both conditions derived from the MS equation and numerical studies suggest a hypersurface of constant scalar field may be the appropriate choice; however, future work should look to clarify a canonical hypersurface for the transition. Furthermore, the joining of $v$ and $v'$ as originally prescribed in the Contaldi approximation has been shown to be insufficient to allow continuity of the first and second fundamental forms describing regions of spacetime separated by a jump in equation of state of the scalar field. 

\begin{figure}
\centering
\includegraphics[]{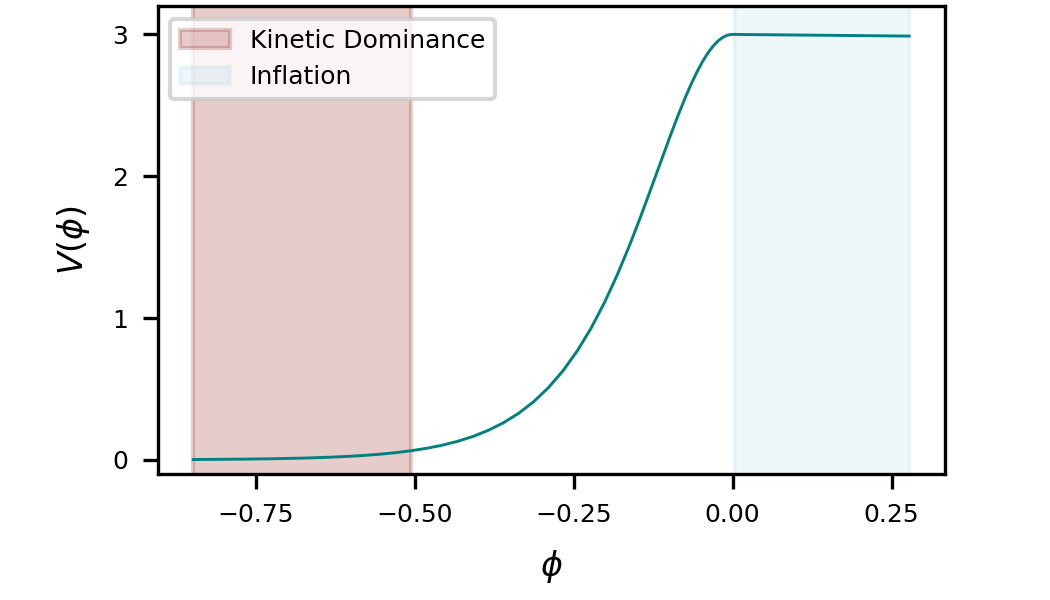}
\caption{Characteristic steplike inflationary potential from parametric reconstruction for a smooth comoving horizon defined by Eq. (83) with a sudden transition corresponding to $\varepsilon_{\mathrm{I}}\ll 1$ and $\alpha>3$. This example has
$\alpha=10$ and $\varepsilon_{\mathrm{I}}=0.0001$. The regions of the potential corresponding to kinetic dominance and inflation are specified using Eqs.  (87) and (88).}
\end{figure}

A novel semianalytic approach for producing the primordial power spectrum, which smoothly transitions from a phase of kinetic dominance to inflation over a finite duration, was subsequently introduced. The difference between primordial power spectra produced from slow and sudden transitions suggests that models describing an instantaneous transition may not adequately characterize primordial power spectra resulting from transitions occurring over several $e$-folds, as is thought to arise in nature. This is supported by the unphysical spectra produced from cosmological matching conditions applied to the Contaldi approximation and the steplike form of the implicit inflationary potential which is demanded to produce sufficiently sudden finite cosmological phase transitions. That is, the alternative model for generating the primordial power spectrum presented in this work whilst it does not require a choice of the form of the inflationary potential, it has an associated potential which can be reconstructed. Further work must be done to constrain primordial power spectra produced from this model and the extended model in Appendix D  
for low $\alpha$ corresponding to a small field implicit inflationary potentials which may be bound observationally via ($n_s$, $r$) \cite{2014A&A...571A..22P, 2020A&A...641A..10P}.

\section{Acknowledgments}

We would like to thank Enrico Pajer and Carlo Contaldi for valuable feedback on this work as well as Thomas Gessey-Jones for many useful conversations.

\appendix

\section{FORMULATION OF THE ISRAEL JUNCTION CONDITIONS}

The Israel junction conditions are formulated as in \citet{poisson_2004}, where necessary background is first introduced. Consider a hypersurface (timelike or spacelike), $\Sigma$, partitioning spacetime into two regions $\mathcal{V}^+$ with metric $g_{\alpha \beta}^+$ expressed with coordinates $x_+^{\alpha}$ and $\mathcal{V}^-$ with metric $g_{\alpha \beta}^-$ expressed with coordinates $x_-^{\alpha}$. The conditions that must be imposed on the metrics to allow the regions of spacetime $\mathcal{V}^+$ and $\mathcal{V}^-$ to join smoothly at $\Sigma$ to allow for the union of the metric to form valid solutions to the Einstein field equations are studied \cite{poisson_2004,2023arXiv230312457L, 2005PhRvD..72d4016M,2020CQGra..37g5022A}.

Assume the coordinates $y^a$ may be installed  on both sides of the hypersurface and select the unit normal to $\Sigma$,  $n^{\alpha}$, to point from $\mathcal{V}^-$ to $\mathcal{V}^+$. There also exists a distinct coordinate system $x^{\alpha}$ that may be installed on both sides of the hypersurface which overlaps with the coordinates defined on either side of the hypersurface in an open region containing $\Sigma$. Imagine $\Sigma$ to be pierced by a congruence of geodesics that intersect it orthogonally. $\ell$ is taken to denote proper distance or time along the geodesics ($\ell=0$ when the geodesic crosses the hypersurface) and should be thought of as a scalar field. The point $P$ denoted by the coordinates $x^{\alpha}$ is linked to $\Sigma$ by a member of the congruence and $\ell(x^{\alpha})$ is the proper distance or time from the hypersurface along the geodesic described by

\begin{equation}
    dx^{\alpha}=n_{\alpha}d\ell,
\end{equation}

\begin{equation}
    n_{\alpha}=\sigma \partial_{\alpha}\ell,
\end{equation}

\noindent
where $n^{\alpha}n_{\alpha}=\sigma$.

The tensor quantity, $A$, exists on both sides of the hypersurface and $[A]_{\pm}$ denotes a jump across the hypersurface. This jump notation is defined as follows. Using coordinates such that the equation of $\Sigma$ is $f(x)=0$ and $\mathcal{V}^+$ is $f(x)>0$ and  $\mathcal{V}^-$ is $f(x)<0$.
For a function $h(x)$ defined on either side of the hypersurface, if $h$ has at most a simple discontinuity at $\Sigma$,

$$[h](P)=\lim_{Q\to P} h^+-\lim_{R\to P}h^-,$$

\noindent
where $P\in \Sigma$ and $Q$ and $R$ tend to $P$ through $\mathcal{V}^+$  and $\mathcal{V}^-$, respectively. For a small displacement, $\delta$, taking $Q=P+\delta$ and $R=P-\delta$ this can be written as

$$[h]_{\pm}=\int^{P+\delta}_{P-\delta}h'dx=h(P+\delta)-h(P-\delta).$$

Using Eq. (A1) and continuity of $\ell$ and $x^{\alpha}$ across the hypersurface gives

\begin{equation}
    [n^{\alpha}]_{\pm}=0.
\end{equation}
\noindent
Recalling that the coordinate $y^{a}$ are the same on both sides of $\Sigma$,

\begin{equation}
    [e^{\alpha}_a]_{\pm}=0.
\end{equation}

\noindent
Here, the the Heaviside distribution is introduced,

\begin{equation}
    \Theta(\ell)= \begin{cases} 
      1 & \ell>0 \\
      0 & \ell<0  \\
   \end{cases},
\end{equation}
\noindent
where if $\ell=0$ the function is indeterminate. It should be noted that the product of the Heaviside distribution and Dirac distribution, $\delta(\ell)$, is not defined as a distribution. As well, the Dirac distribution, $\delta(\ell)$, is not be confused with the small displacement, $\delta$, used above to define the jump notation.

Looking at the first Israel junction condition, the metric, $g_{\alpha \beta}$, in coordinates $x^{\alpha}$ as a distribution-valued tensor is 

\begin{equation}
g_{\alpha \beta}=\Theta(\ell)g_{\alpha \beta}^{+}+\Theta(-\ell)g_{\alpha\beta}^{-},
\end{equation}
\noindent
where $g_{\alpha\beta}^{\pm}$ is the metric in $\mathcal{V}^{\pm}$ expressed in the coordinates $x^{\alpha}$. Whether Eq. (A6) makes a valid distributional solution to the Einstein equations must be considered by verifying geometrical quantities constructed from the metric defined by this equation. To do this requires differentiating, which yields

\begin{equation}
g_{\alpha\beta, \gamma}=\Theta(\ell)g_{\alpha\beta, \gamma}^{+}+\Theta(-\ell)g_{\alpha\beta, \gamma}^{-}+\sigma\delta(\ell)[g_{\alpha\beta}]_{\pm}n_{\gamma}.
\end{equation}
\noindent
The last term is singular and causes problems when computing the Christoffel symbols as it generates terms proportional to $\Theta(\ell)\delta(\ell)$, which would make the connection not a valid distribution. Eliminating the last term in Eq. (A7) imposes continuity of the metric across the hypersurface through the following condition:

\begin{equation}
     \big[g_{\alpha\beta}\big]_{\pm}=0.
\end{equation}

\noindent
As a coordinate invariant statement, by Eq. (A4),

\begin{equation}
     \big[g_{\alpha\beta}e^{\alpha}_ae^{\beta}_b\big]_{\pm}=0,
\end{equation}

\noindent
and the first junction condition may be interpreted as continuity of the induced metric on either side of the hypersurface,

\begin{equation}
    \big[h_{ab}\big]_{\pm}=0.
\end{equation}

The second Israel junction condition requires a look at the distribution-valued Riemann tensor. The Christoffel symbols are 

\begin{equation}
\Gamma_{\beta\gamma}^{\alpha}=\Theta(\ell)\Gamma_{ \beta\gamma}^{+\alpha}+\Theta(-\ell)\Gamma_{\beta\gamma}^{-\alpha}.
\end{equation}

\noindent
Differentiating the above equation, the Riemann tensor follows as

\begin{equation}
R_{\beta\gamma\delta}^{\alpha}=\Theta(\ell)R_{ \beta\gamma\delta}^{+\alpha}+\Theta(-\ell)R_{\beta\gamma\delta}^{-\alpha}+\delta(\ell)A^{\alpha}_{\beta\gamma\delta},
\end{equation}
\noindent
where 

\begin{equation}
A_{\beta\gamma\delta}^{\alpha}=\sigma\Big([\Gamma_{ \beta\delta}^{\alpha}]_{\pm}n_{\gamma}-[\Gamma_{ \beta\gamma}^{\alpha}]_{\pm}n_{\delta}\Big).
\end{equation}

\noindent
The second junction condition aims to  eliminate the final term in Eq. (A12) as although it is properly defined as a distribution, the $\delta(\ell)$ function term represents a curvature singularity at $\Sigma$. The first junction condition implies continuity of the metric across $\Sigma$ in the coordinates $x^{\alpha}$ and requires tangential derivatives to also be continuous. Therefore, if $g_{\alpha \beta ,\gamma}$ is discontinuous it must be along the normal vector $n^{\alpha}$, and there must exist a tensor field, $\kappa_{\alpha \beta}$ such that 

\begin{equation}
    [g_{\alpha \beta ,\gamma}]_{\pm}=\kappa_{\alpha \beta}n_{\gamma},
\end{equation}

\noindent
with

\begin{equation}
\kappa_{\alpha \beta}=\sigma    [g_{\alpha \beta ,\gamma}]_{\pm}n^{\gamma}.
\end{equation}

\noindent
Equation (A15) implies 

$$[\Gamma^{\alpha}_{\beta\gamma}]_{\pm}=\frac{1}{2}(\kappa^{\alpha}_{\beta}n_{\gamma}+\kappa^{\alpha}_{\gamma}n_{\beta}-\kappa_{\beta\gamma}n^{\alpha}).$$

\noindent 
Then the $\delta(\ell)$ function part of the Riemann tensor is 

$$A^{\alpha}_{\beta\gamma\delta}=\frac{\sigma}{2}(\kappa^{\alpha}_{\delta}n_{\beta}n_{\gamma}+
\kappa^{\alpha}_{\gamma}n_{\beta}n_{\delta}-\kappa_{\beta\delta}n^{\alpha}n_{\gamma}+\kappa_{\beta\gamma}n_{\alpha}n_{\delta}).$$

\noindent 
 One can obtain the distribution-valued function for the stress-energy given in the following equation by contracting indices twice to attain the $\delta(\ell)$ function part of the Ricci scalar. The stress-energy tensor is given by

\begin{equation}
    T_{\alpha\beta}=\Theta(\ell)T_{\alpha\beta}^{+}+\Theta(-\ell)T_{\alpha\beta}^{-}+\delta(\ell)S_{\alpha\beta},
\end{equation}

\noindent
where the surface stress-energy surface layer, $S_{ab}$, is given by

\begin{multline}
    16\pi \sigma S_{ab}=\kappa_{\mu\alpha}n^{\mu}n_{\beta}+\kappa_{\mu\beta}n^{\mu}n_{\alpha}-\kappa n_{\alpha}n_{\beta}\\
    -\sigma\kappa_{\alpha\beta}-(\kappa_{\mu\nu} n^{\mu}n^{\nu}-\sigma\kappa)g_{\alpha\beta}.
\end{multline}

\noindent
Through careful calculation detailed in Ref. \cite{poisson_2004}, one arrives at an expression for the surface stress-energy tensor in terms of a jump in extrinsic curvature, $K_{ab}$,

\begin{equation}
    S_{ab}=-\frac{\sigma}{8\pi}\Big([K_{ab}]_{\pm}-[K]_{\pm}h_{ab}\Big),
\end{equation}
\noindent
and for a smooth transition across $\Sigma$, the second junction condition is concluded as requiring the extrinsic curvature on either side of the hypersurface to be the same. This is

\begin{equation}
    \big[K_{ab}\big]_{\pm}=0.
\end{equation}
\noindent

In the absence of a surface stress-energy tensor, the Israel junction conditions are then given by Eqs.  (A10) and (A19), which demand continuity of the induced metric and extrinsic curvature for a smooth joining of two metrics.

\section{SCALAR METRIC RELATIONS}

Equation (60) is derived using  \citet{1992PhR...215..203M}.
Working in the Newtonian/longitudinal gauge, expression (56a) describes cosmological perturbations for a universe dominated by a scalar field,

$$\Phi'+\mathcal{H}\Phi=\frac{1}{2} \delta \phi \bar{\phi}'. $$

\noindent
Varying the action for a scalar field with respect to $\phi$ and $\psi$ and setting $B-E'=0$ gives the equation, 

\begin{equation}
    \Delta\Psi-3\mathcal{H}\Psi'-(\mathcal{H}'+2\mathcal{H}^2)\Phi=\frac{1}{2}(\delta \phi \bar{\phi}'+V_{,\phi}a^2\delta \phi).
\end{equation}

\noindent
Recall the  definition of $\mathcal{R}$ from Eq. (57),

$$\mathcal{R}=\Phi-\frac{\mathcal{H}}{\mathcal{H}^2-\mathcal{H}'}(\Phi'+\mathcal{H}\Phi).$$

\noindent
Using Eqs.  (56a), (B1), and (57) as well as noting that in the Newtonian/longitudinal gauge $\Phi=\Psi$, Eq. (60) is recovered 

\begin{equation}
    \Delta \Phi=\frac{1}{2}\frac{\bar{\phi}'^2}{\mathcal{H}}\mathcal{R}'.
\end{equation}

Next, the relationship in Eq. (62) is derived following \citet{2012JCAP...12..018N}. The gauge-invariant Bardeen potentials are 

\begin{equation}
    \Phi_{\mathrm{B}}=\Phi-\dot{\theta},
\end{equation}

\begin{equation}
    \Psi_{\mathrm{B}}=\Psi+H\theta,
\end{equation}

\noindent
where $\theta=a^2[\dot{E}-B/a]$. Working in an isotropic background with no anisotropic stress $\Psi_{\mathrm{B}}=\Phi_{\mathrm{B}}$. In the comoving gauge ($\delta q=0 $, $E=0$), the comoving curvature perturbation is 

\begin{equation}
    \mathcal{R}=\Psi.
\end{equation}

\noindent
The $0i$ components of the Einstein equations in the comoving gauge give

\begin{equation}
    \dot{\mathcal{R}}+H\Phi=0.
\end{equation}

\noindent
Differentiating, we have the relation, 

\begin{equation}
    \ddot{\mathcal{R}}=-H\dot{\Phi}-\dot{H}\Phi.
\end{equation}

\noindent
In the absence of anisotropic stress, the $00$ and $ii$ components of the Einstein equations in the comoving gauge are

\begin{equation}
    3H(\dot{\mathcal{R}}+H\Phi)+\frac{k^2}{a^2}(aHB-\mathcal{R})=\frac{\delta \rho}{2},
\end{equation}

\begin{equation}
-\ddot{\mathcal{R}}-3H\dot{\mathcal{R}}-H\dot{\Phi}+(2\dot{H}-3H^2)\Phi=\frac{\delta p}{2}.
\end{equation}

\noindent
Using Eqs.  (B6) and (B7) to simplify (B8) and (B9) gives

\begin{equation}
    \frac{k^2}{a^2}(aHB-\mathcal{R})=\frac{\delta \rho}{2},
\end{equation}

\begin{equation}
\frac{\dot{H}}{H}\dot{R}=\frac{\delta p}{2}.
\end{equation}

\noindent
The sound speed of perturbations, $c_s$, is defined on comoving slices as $\delta p_c=c_s^2\delta \rho_c$, where $\delta p_c$ and $\delta \rho_c$ are the pressure and energy density perturbations on the comoving slices. Using Eqs.  (B10) and (B11),

\begin{equation}
    \Phi_{\mathrm{B}}=-\frac{a^2\dot{H}}{Hk^2c_s^2}\dot{\mathcal{R}}.
\end{equation}

\noindent
Changing to conformal time, recalling the definition of $z$ and taking $c_s=1$, the gauge-invariant potential is 

\begin{equation}
    \Phi_{\mathrm{B}}=\frac{z^2H}{2ak^2}\mathcal{R}'.
\end{equation}

\noindent 
Switching to the Newtonian/longitudinal gauge, $\Phi=\Phi_{\mathrm{B}}$, Eq. (62) is obtained

\begin{equation}
    \Phi=\frac{z^2H}{2ak^2}\mathcal{R}'.
\end{equation}

\section{CONSTANTS OF INTEGRATION}

Solving $A_k$ and $B_k$ for each vacuum initial conditions outlined in Table 1 and imposing Contaldi matching, Eqs.  (32)-(33), the constants of integration are

\noindent
For right-handed mode (RHM),

\begin{multline}
     C^{\mathrm{(RHM)}}_k=\sqrt{\frac{\pi}{32k_{\mathrm{t}}}}e^{-i\frac{k}{k_{\mathrm{t}}}}\Bigg(H_0^{(2)}\Big[\frac{k}{2k_{\mathrm{t}}}\Big]\\
     -\Big(\frac{k_{\mathrm{t}}}{k}+i\Big)H_1^{(2)}\Big[\frac{k}{2k_{\mathrm{t}}}\Big]\Bigg),
\end{multline}

\begin{multline}
    D^{\mathrm{(RHM)}}_k=\sqrt{\frac{\pi}{32k_{\mathrm{t}}}}e^{+i\frac{k}{k_{\mathrm{t}}}}\Bigg(H_0^{(2)}\Big[\frac{k}{2k_{\mathrm{t}}}\Big]\\
    -\Big(\frac{k_{\mathrm{t}}}{k}-i\Big)H_1^{(2)}\Big[\frac{k}{2k_{\mathrm{t}}}\Big]\Bigg).
\end{multline}

\noindent
For Bunch-Davies vacuum (BD),

\begin{equation}
    C^{\mathrm{(BD)}}_k=\sqrt{\frac{1}{8k}}e^{-i\frac{k}{k_{\mathrm{t}}}}\Bigg(\Big(\frac{k_{\mathrm{t}}}{k}\Big)^2+2i\Big(\frac{k_{\mathrm{t}}}{k}\Big)-2\Bigg),
\end{equation}

\begin{equation}
    D^{\mathrm{(BD)}}_k=\sqrt{\frac{1}{8k}}e^{+i\frac{k}{k_{\mathrm{t}}}}\Big(\frac{k_{\mathrm{t}}}{k}\Big)^2.
\end{equation}

\noindent
For Hamiltonian diagonalization (HD),

\begin{multline}
    C^{\mathrm{(HD)}}_k=\sqrt{\frac{1}{8k}}e^{-i\frac{k}{k_{\mathrm{t}}}}\Bigg(\Big(\frac{k_{\mathrm{t}}}{k}\Big)^2\\
    +i\Big(1+\frac{\omega_k}{k}\Big)\frac{k_{\mathrm{t}}}{k}
    -\Big(1+\frac{\omega_k}{k}\Big)\Bigg),
\end{multline}

\begin{multline}
     D^{\mathrm{(HD)}}_k=\sqrt{\frac{1}{8k}}e^{+i\frac{k}{k_{\mathrm{t}}}}\Bigg(\Big(\frac{k_{\mathrm{t}}}{k}\Big)^2\\
     -i\Big(1-\frac{\omega_k}{k}\Big)\frac{k_{\mathrm{t}}}{k}-\Big(1-\frac{\omega_k}{k}\Big)\Bigg),
\end{multline}
\noindent
where $\omega_k=k^2+k_{\mathrm{t}}^2$. For renormalized stress energy tensor (RSET),

\begin{equation}
    C^{\mathrm{(RSET)}}_k=\sqrt{\frac{1}{8k}}e^{-i\frac{k}{k_{\mathrm{t}}}}\Big(i\frac{k_{\mathrm{t}}}{k}-2\Big),
\end{equation}

\begin{equation}
    D^{\mathrm{(RSET)}}_k=\sqrt{\frac{1}{8k}}e^{+i\frac{k}{k_{\mathrm{t}}}}\Big(i\frac{k_{\mathrm{t}}}{k}\Big).
\end{equation}

For a hypersurface defined by $\Sigma: \bar{\rho}+\delta \rho=\text{const.}$ with matching conditions (63a)-(63b). The constants of integration for the vacuum conditions in Table 1 are 

\noindent
For right-handed mode (RHM),

\begin{multline}
        C^{\mathrm{(RHM)}}_k=\sqrt{\frac{\pi}{864k_{\mathrm{t}}\varepsilon_{\mathrm{I}}}}e^{-i\frac{k}{k_{\mathrm{t}}}}\Bigg(-3\varepsilon_{\mathrm{I}}H_0^{(2)}\Big[\frac{k}{2k_{\mathrm{t}}}\Big]\\
    +\Big(-3\frac{k}{k_{\mathrm{t}}}+9i+9\frac{k_{\mathrm{t}}}{k}+\varepsilon_{\mathrm{I}}\frac{k}{k_{\mathrm{t}}}\Big)H_1^{(2)}\Big[\frac{k}{2k_{\mathrm{t}}}\Big]\Bigg),
\end{multline}

\begin{multline}
    D^{\mathrm{(RHM)}}_k=\sqrt{\frac{\pi}{864k_{\mathrm{t}}\varepsilon_{\mathrm{I}}}}e^{+i\frac{k}{k_{\mathrm{t}}}}\Bigg(-3\varepsilon_{\mathrm{I}}H_0^{(2)}\Big[\frac{k}{2k_{\mathrm{t}}}\Big]\\
    +\Big(-3\frac{k}{k_{\mathrm{t}}}-9i+9\frac{k_{\mathrm{t}}}{k}+\varepsilon_{\mathrm{I}}\frac{k}{k_{\mathrm{t}}}\Big)H_1^{(2)}\Big[\frac{k}{2k_{\mathrm{t}}}\Big]\Bigg).
\end{multline}

\noindent
For Bunch-Davies vacuum (BD),

\begin{multline}
    C^{\mathrm{(BD)}}_k=\sqrt{\frac{1}{216k\varepsilon_{\mathrm{I}}}}e^{-i\frac{k}{k_{\mathrm{t}}}}\Bigg(-3i\frac{k}{k_{\mathrm{t}}}-12\\
    +18i\frac{k_{\mathrm{t}}}{k}+9\Big(\frac{k_{\mathrm{t}}}{k}\Big)^2+i\varepsilon_{\mathrm{I}}\frac{k}{k_{\mathrm{t}}}-2\varepsilon_{\mathrm{I}}\Bigg),
\end{multline}

\begin{multline}
   D^{\mathrm{(BD)}}_k=\sqrt{\frac{1}{216k\varepsilon_{\mathrm{I}}}}e^{+i\frac{k}{k_{\mathrm{t}}}}\Bigg(-3i\frac{k}{k_{\mathrm{t}}}+6+9\Big(\frac{k_{\mathrm{t}}}{k}\Big)^2\\
   +i\varepsilon_{\mathrm{I}}\frac{k}{k_{\mathrm{t}}}-2\varepsilon_{\mathrm{I}}\Bigg).
\end{multline}

\noindent
For Hamiltonian diagonalization (HD),

\begin{multline}
    C^{\mathrm{(HD)}}_k=\sqrt{\frac{1}{216k\varepsilon_{\mathrm{I}}}}e^{-i\frac{k}{k_{\mathrm{t}}}}\Bigg(3-9i\frac{k_{\mathrm{t}}}{k}-9\Big(\frac{k_{\mathrm{t}}}{k}\Big)^2\\
    +3i\frac{\omega_k}{k_{\mathrm{t}}}
    +9\frac{\omega_k}{k}-9i\frac{k_{\mathrm{t}}}{k}\frac{\omega_k}{k}-i\varepsilon_{\mathrm{I}}\frac{\omega_k}{k_{\mathrm{t}}}+2\varepsilon_{\mathrm{I}}\Bigg),
\end{multline}

\begin{multline}
    D^{\mathrm{(HD)}}_k=\sqrt{\frac{1}{216k\varepsilon_{\mathrm{I}}}}e^{+i\frac{k}{k_{\mathrm{t}}}}\Bigg(3+9i\frac{k_{\mathrm{t}}}{k}-9\Big(\frac{k_{\mathrm{t}}}{k}\Big)^2\\
    +3i\frac{\omega_k}{k_{\mathrm{t}}}
    -9\frac{\omega_k}{k}-9i\frac{k_{\mathrm{t}}}{k}\frac{\omega_k}{k}-i\varepsilon_{\mathrm{I}}\frac{\omega_k}{k_{\mathrm{t}}}+2\varepsilon_{\mathrm{I}}\Bigg),
\end{multline}

\noindent
where $\omega_k=k^2+k_{\mathrm{t}}^2$. For renormalized stress energy tensor (RSET),

\begin{equation}
    C^{\mathrm{(RSET)}}_k=\sqrt{\frac{1}{216k\varepsilon_{\mathrm{I}}}}e^{-i\frac{k}{k_{\mathrm{t}}}}\Bigg(3i\frac{k}{k_{\mathrm{t}}}+9-9i\frac{k_{\mathrm{t}}}{k}-i\varepsilon_{\mathrm{I}}\frac{k}{k_{\mathrm{t}}}+3\varepsilon_{\mathrm{I}}\Bigg),
\end{equation}

\begin{equation}
     D^{\mathrm{(RSET)}}_k=\sqrt{\frac{1}{216k\varepsilon_{\mathrm{I}}}}e^{+i\frac{k}{k_{\mathrm{t}}}}\Bigg(3i\frac{k}{k_{\mathrm{t}}}-9-9i\frac{k_{\mathrm{t}}}{k}-i\varepsilon_{\mathrm{I}}\frac{k}{k_{\mathrm{t}}}+3\varepsilon_{\mathrm{I}}\Bigg).
\end{equation}

For a hypersurface defined by $\Sigma:\bar{\phi}+\delta \phi=\text{const.}$ with matching conditions (68a)-(68b). The constants of integration for the vacuum conditions in Table 1 are

\noindent
For right-handed mode (RHM),

\begin{multline}
C^{\mathrm{(RHM)}}_k=\sqrt{\frac{\pi}{96k_{\mathrm{t}}\varepsilon_{\mathrm{I}}}}e^{-i\frac{k}{k_{\mathrm{t}}}}\Bigg(\varepsilon_{\mathrm{I}}H_0^{(2)}\Big[\frac{k}{2k_{\mathrm{t}}}\Big]\\
    -3\Big(\frac{k_{\mathrm{t}}}{k}+i\Big)H_1^{(2)}\Big[\frac{k}{2k_{\mathrm{t}}}\Big]\Bigg),
\end{multline}

\begin{multline}
    D^{\mathrm{(RHM)}}_k=\sqrt{\frac{\pi}{96k_{\mathrm{t}}\varepsilon_{\mathrm{I}}}}e^{+i\frac{k}{k_{\mathrm{t}}}}\Bigg(\varepsilon_{\mathrm{I}}H_0^{(2)}\Big[\frac{k}{2k_{\mathrm{t}}}\Big]\\
    -3\Big(\frac{k_{\mathrm{t}}}{k}-i\Big)H_1^{(2)}\Big[\frac{k}{2k_{\mathrm{t}}}\Big]\Bigg).
\end{multline}

\noindent
For Bunch-Davies vacuum (BD),

\begin{multline}
    C^{\mathrm{(BD)}}_k=\sqrt{\frac{1}{24k\varepsilon_{\mathrm{I}}}}e^{-i\frac{k}{k_{\mathrm{t}}}}\Bigg(3\Big(1-2i\frac{k_{\mathrm{t}}}{k}\\
    -\Big(\frac{k_{\mathrm{t}}}{k}\Big)^2\Big)+\varepsilon_{\mathrm{I}}\Bigg),
\end{multline}

\begin{multline}
    D^{\mathrm{(BD)}}_k=\sqrt{\frac{1}{24k\varepsilon_{\mathrm{I}}}}e^{+i\frac{k}{k_{\mathrm{t}}}}\Bigg(-3\Big(1+\Big(\frac{k_{\mathrm{t}}}{k}\Big)^2\Big)+\varepsilon_{\mathrm{I}}\Bigg).
\end{multline}

\noindent
For Hamiltonian diagonalization (HD),

\begin{multline}
    C^{\mathrm{(HD)}}_k=\sqrt{\frac{1}{24k\varepsilon_{\mathrm{I}}}}e^{-i\frac{k}{k_{\mathrm{t}}}}\Bigg(3\bigg(-i\frac{k_{\mathrm{t}}}{k}-\Big(\frac{k_{\mathrm{t}}}{k}\Big)^2\\
    +\frac{\omega_k}{k}-i\frac{k_{\mathrm{t}}}{k}\frac{\omega_k}{k}\bigg)+\varepsilon_{\mathrm{I}}\Bigg),
\end{multline}

\begin{multline}
    D^{\mathrm{(HD)}}_k=\sqrt{\frac{1}{24k\varepsilon_{\mathrm{I}}}}e^{+i\frac{k}{k_{\mathrm{t}}}}\Bigg(3\bigg(i\frac{k_{\mathrm{t}}}{k}-\Big(\frac{k_{\mathrm{t}}}{k}\Big)^2\\
    -\frac{\omega_k}{k}-i\frac{k_{\mathrm{t}}}{k}\frac{\omega_k}{k}\bigg)+\varepsilon_{\mathrm{I}}\Bigg),
\end{multline}

\noindent
where $\omega_k=k^2+k_{\mathrm{t}}^2$. For renormalized stress energy tensor (RSET),

\begin{equation}
    C^{\mathrm{(RSET)}}_k=\sqrt{\frac{1}{24k\varepsilon_{\mathrm{I}}}}e^{-i\frac{k}{k_{\mathrm{t}}}}\Bigg(3\Big(1-i\frac{k_{\mathrm{t}}}{k}\Big)+\varepsilon_{\mathrm{I}}\Bigg),
\end{equation}

\begin{equation}
    D^{\mathrm{(RSET)}}_k=\sqrt{\frac{1}{24k\varepsilon_{\mathrm{I}}}}e^{+i\frac{k}{k_{\mathrm{t}}}}\Bigg(-3\Big(1+i\frac{k_{\mathrm{t}}}{k}\Big)+\varepsilon_{\mathrm{I}}\Bigg).
\end{equation}

\section{CONSTRUCTING A SOLUTION WITH SLOW-ROLL INFLATION}
 The following procedure will be used to obtain background equations with a slowly increasing $\varepsilon(N)$:

\begin{equation}
   \text{Modify } \frac{\varepsilon'}{\varepsilon}\xrightarrow{\mathrm{solve}}\varepsilon=-\frac{H'}{H}\xrightarrow{\mathrm{solve}}H\xrightarrow{}\frac{1}{aH},
\end{equation}

\noindent
giving a comoving horizon which transitions from kinetic dominance to a slow-roll inflationary phase described by $\varepsilon(N), |\eta(N)|\ll 1$. Beginning with Eq. (80) for $\varepsilon(N)$ for a model with a phase of modified de Sitter inflation,

\begin{equation}
   \frac{\varepsilon'(N)}{\varepsilon(N)}=\frac{3\alpha e^{3\alpha N}(\varepsilon_{\mathrm{I}}-3)}{(1+e^{3\alpha N})(3+\varepsilon_{\mathrm{I}}e^{3\alpha N})},
\end{equation}

\noindent
with

\begin{equation}
    \lim_{N\to-\infty} \frac{\varepsilon'(N)}{\varepsilon(N)}=0,
\end{equation}

\begin{equation}
       \lim_{N\to\infty} \frac{\varepsilon'(N)}{\varepsilon(N)}= 0.
   \end{equation}

\noindent
Then, the limits of the second-slow roll parameter are

\begin{equation}
    \lim_{N\to-\infty} \eta(N)=\varepsilon_{\mathrm{KD}},
\end{equation}

\begin{equation}
    \lim_{N\to\infty} \eta(N)=\varepsilon_{\mathrm{I}}.
\end{equation}

\noindent
Modifying the Eq. (D2) to be

\begin{equation}
  \frac{\varepsilon'(N)}{\varepsilon(N)}=\frac{3\alpha e^{3\alpha N}(\varepsilon_{\mathrm{I}}-3)-3C}{(1+e^{3\alpha N})(3+\varepsilon_{\mathrm{I}}e^{3\alpha N})}+C,
\end{equation}

\noindent
will allow for a time varying $\varepsilon(N)$ parametrized by C. The solution to Eq. (D7) gives the following equation: 

\begin{equation}
    \varepsilon(N)=(1+e^{3\alpha N})^{-1+\frac{C}{3\alpha-\alpha \varepsilon_{\mathrm{I}}}}(3+\varepsilon_{\mathrm{I}}e^{3\alpha N})^{1+\frac{C\varepsilon_{\mathrm{I}}}{3\alpha(\varepsilon_{\mathrm{I}}-3)}},
\end{equation}

\noindent
with inflation ending when the slow-roll conditions are violated, that is at $\varepsilon(N_{\mathrm{end}})=1$ \cite{2009arXiv0907.5424B}. The relevant limits of Eq. (D8) are

\begin{equation}
    \lim_{N\to-\infty} \frac{\varepsilon'(N)}{\varepsilon(N)}=0,
\end{equation}

\begin{equation}
       \lim_{N\to N_{\mathrm{end}}} \frac{\varepsilon'(N)}{\varepsilon(N)}=C.
   \end{equation}

\noindent
The limits of the second-slow roll parameter become

\begin{equation}
    \lim_{N\to-\infty} \eta(N)=\varepsilon_{\mathrm{KD}},
\end{equation}

\begin{equation}
    \lim_{N\to N_{\mathrm{end}}} \eta(N)=1-\frac{C}{2}.
\end{equation}

\noindent
Using Eq. (D8), the background equations are

\begin{equation}
    \frac{H'(N)}{H(N)}=-(1+e^{3\alpha N})^{-1+\frac{C}{3\alpha-\alpha \varepsilon_{\mathrm{I}}}}(3+\varepsilon_{\mathrm{I}}e^{3\alpha N})^{1+\frac{C\varepsilon_{\mathrm{I}}}{3\alpha(\varepsilon_{\mathrm{I}}-3)}},
\end{equation}

\begin{multline}
    \frac{z'(N)}{z(N)}=\frac{6}{2(1+e^{3\alpha N})(3+\varepsilon_{\mathrm{I}}e^{3\alpha N})}\\
    +\frac{e^{3 \alpha N}\big(3\alpha(\varepsilon_{\mathrm{I}}-3)+(2+C)(3+\varepsilon_{\mathrm{I}}+\varepsilon_{\mathrm{I}}e^{3\alpha N})\big)}{2(1+e^{3\alpha N})(3+\varepsilon_{\mathrm{I}}e^{3\alpha N})}.
\end{multline}

\noindent
The analytic equation for $\frac{1}{a(N)H(N)}$ may be found by solving Eq. (D14) in terms of $H(N)$ and can be written in terms of Appell series and hypergeometric functions with parameters $\varepsilon_{\mathrm{I}}$, $\alpha$, and $C$ . Using these analytic equations for the background, the MS Eq. (72) can be solved numerically in order to produce a primordial power spectrum.

\bibliography{references}

\end{document}